\documentclass[a4paper,12pt,reqno]{article}

\usepackage{amsmath,amsfonts,amssymb,amsthm,dsfont,slashed}
\allowdisplaybreaks

\newcommand{\da}{{\dot \alpha}}
\newcommand{\dbe}{{\dot \beta}}
\newcommand{\dga}{{\dot \gamma}}

\newcommand{\cg}{{\cal G}}
\newcommand{\ch}{{\cal H}}

\newcommand{\cD}{{\cal D}}
\newcommand{\cP}{{\cal P}}

\newcommand{\cT}{{\cal T}}
\newcommand{\cU}{{\cal U}}
\newcommand{\cV}{{\cal V}}
\newcommand{\qd}{{\quad}}

\newcommand{\f}[3]{{f_{#1#2}}{}^{#3}}

\newcommand{\viel}[2]{e_{#2}{}^{#1}}
\newcommand{\Viel}[2]{e^{#2}{}_{#1}}
\newcommand{\A}[2]{A_{#1}{}^{#2}}
\newcommand{\De}[2]{\delta^{(#1)}_{#2}}
\newcommand{\F}[2]{F_{#1}{}^{#2}}
\newcommand{\T}[2]{T_{#1}{}^{#2}}
\newcommand{\EL}[2]{\frac{\delta #1}{\delta #2}}

\newcommand{\bea}{\begin{eqnarray}} \newcommand{\eea}{\end{eqnarray}}
\newcommand{\beann}{\begin{eqnarray*}} \newcommand{\eeann}{\end{eqnarray*}}
\newcommand{\beq}{\begin{equation}} \newcommand{\eeq}{\end{equation}}
\newcommand{\ba}{\begin{array}} \newcommand{\ea}{\end{array}}
\newcommand{\ben}{\begin{enumerate}} \newcommand{\een}{\end{enumerate}}

\newcommand{\Ii}{\mathrm{i}}

\newcommand{\4 }{\tilde}
\newcommand{\5}{\bar}
\newcommand{\6}{\partial}

\newcommand{\ul}[1]{{\underline{#1}}}

\newcommand{\com}[2]{[\,#1\, ,\,#2\,]}	
\newcommand{\acom}[2]{\{#1\, ,\,#2\}}	

\newcommand{\csum}[2]{\sum_{#1}\hspace{-1.#2em}\circ\ \ \ }

\newcommand{\LB}{L^{(0)}_\mathrm{Bose}}
\newcommand{\LBo}{L^{(1)}_\mathrm{Bose}}
\newcommand{\LF}{L^{(0)}_\mathrm{Fermi}}
\newcommand{\LS}{L^{(0)}_\mathrm{susy}}

\newcommand{\mysection}[1]{\section{#1}
           \setcounter{equation}{0}\setcounter{figure}{0}}

\begin{document}

\begin{center}
 {\large\bfseries 4D supersymmetric gauge theories of spacetime translations}
 \\[5mm]
 Friedemann Brandt \\[2mm]
 \textit{Institut f\"ur Theoretische Physik, Leibniz Universit\"at Hannover, Appelstra\ss e 2, 30167 Hannover, Germany}
\end{center}

\begin{abstract}
The paper addresses the question whether in four spacetime dimensions, besides standard supergravity theories, field theories exist
whose symmetries include local spacetime translations 
and supersymmetries generated by transformations whose commutators contain infinitesimal local spacetime translations. It is shown that, up to local field redefinitions, there is a unique class of such supersymmetric field theories satisfying specified prerequisites which particularly concern the field content of the theories. The theories of this class have global supersymmetry and are similar to globally supersymmetric Yang-Mills theories.
\end{abstract}

\mysection{Introduction}

According to its title, the present paper concerns supersymmetric gauge theories of spacetime translations in four spacetime dimensions. We define such theories as field theories whose Lagrangians are invariant, up to total divergences, at least under local spacetime translations and global supersymmetry transformations. Local spacetime translations of a field $\phi$ are generated by infinitesimal transformations $\delta\phi=v^\mu\6_\mu\phi+\dots$ where the $v^\mu$ are gauge parameters, i.e. the $v^\mu$ are arbitrary functions $v^\mu(x)$ of the spacetime coordinates $x^\mu$ (the ellipses denote possible additional terms which generally depend on the particular field $\phi$). The infinitesimal global supersymmetry transformations are required to have a standard commutator algebra involving infinitesimal spacetime translations (with generally field dependent gauge parameters).

Of course, supersymmetric  gauge theories of spacetime translations are well-known: these are standard supergravity theories \cite{Freedman:1976xh,Deser:1976eh}, cf. \cite{VanNieuwenhuizen:1981ae} for a review, where the local spacetime translations are general coordinate transformations and the supersymmetries are local (= gauge) symmetries. The present paper addresses the question whether there are supersymmetric gauge theories of spacetime translations which are different from standard supergravity theories.  

Concretely we seek globally supersymmetric extensions of actions of theories which sometimes are called ``teleparallel'' theories (henceforth this term will be used for these theories -- just to name them). The Lagrangians of these theories contain linear combinations of terms which are quadratic in the components of a torsion constructed out of the spacetime derivatives of a tetrad field. 

Section \ref{nstele} briefly reviews these ``teleparallel'' theories as consistent deformations \cite{Barnich:1993vg,Henneaux:1997bm}
of free field theories. In particular it is pointed out that and how standard general coordinate transformations arise as local spacetime translations by consistently deforming these free theories.

In section \ref{susyfree} supersymmetric extensions of the free field theories underlying the ``teleparallel'' theories given in section \ref{nstele} are derived. 

Section \ref{Lnl} outlines how supersymmetric free field theories derived in section \ref{susyfree} may be extended to supersymmetric theories with interacting fields which are supersymmetric extensions of ``teleparallel'' theories as in section \ref{nstele}, and of generalizations of such theories with couplings to other supersymmetry multiplets (super Yang-Mills multiplets, scalar multiplets) and/or higher powers in the torsion and/or terms with more derivatives. Furthermore one class of such theories is worked out explicitly.

Section \ref{add} presents results of an addendum to the published version of the paper which concerns the question to which extent supersymmetric field theories described in sections \ref{susyfree} and \ref{Lnl} are equivalent, respectively.

Section \ref{conc} briefly summarizes results derived in the paper.

Conventions used in the paper are given in appendix \ref{conv}. In addition we use a parameter $z\in\{1,-1,\Ii,-\Ii\}$ which allows one to adapt the results to various conventions used in the literature concerning complex conjugation and signs in the commutator algebra of infinitesimal supersymmetry transformations. Using $z$, our convention concerning complex conjugation is
\begin{align} 
\overline{XY}=z^{2|X|\, | Y|}\5X\ \5Y,\qd z\in\{1,-1,\Ii,-\Ii\}
\end{align}
where $\overline{XY}$ denotes the complex conjugate of the product $XY$ of any two objects $X$ and $Y$, such as fields, operators or differential forms, $\5X$ and $\5Y$ denote the complex conjugates of these objects, and $|X|$ and $|Y|$ denote the Gra\ss mann grading of $X$ and $Y$, respectively, with $|X|,|Y|\in\{0,1\}$ (the Gra\ss mann grading of a ``bosonic'' object is 0, the Gra\ss mann grading of a ``fermionic'' object is 1). Owing to these conventions, $z$ occurs in various equations throughout the paper.

Appendix \ref{algebra} summarizes the derivation of results on the supersymmetry transformations used in section \ref{susyfree}.

Appendix \ref{brst} outlines the derivation of results on the construction of supersymmetric actions used in section \ref{Lnl}.

\mysection{Non-supersymmetric ``teleparallel'' gauge theories of spacetime translations}\label{nstele}

In this section it is shown how, along the lines of \cite{Brandt:2001hs}, non-supersymmetric gauge theories of spacetime translations can be constructed as consistent deformations of free field theories for a set of abelian gauge fields $\A\mu\nu$ with Lorentz vector indices $\mu,\nu$ in flat four-dimensional spacetime and the abelian gauge transformations 
\begin{align}
A^\prime_\mu{}^\nu=\A\mu\nu+\De0v\A\mu\nu,\qd
\De0v\A\mu\nu=\6_\mu v^\nu
\label{a1}
\end{align}
where the $v^\nu$ are gauge parameters depending arbitrarily on the spacetime coordinates. Our starting point is a free field theory with the Lagrangian  
\begin{align}
\LB=
a_1 F_{\mu\nu\rho}F^{\mu\nu\rho}
+a_2F_{\mu\nu\rho}F^{\rho\mu\nu}
+a_3\F{\mu\nu}\nu F^{\mu\rho}{}_\rho
+a_4\epsilon^{\mu\nu\rho\sigma}F_{\mu\nu\rho}\F{\sigma\tau}\tau
\label{a2}
\end{align}
where $a_1,\dots,a_4$ are (so far arbitrary) real coefficients, $\F{\mu\nu}\rho$ are the components of gauge invariant field strengths
\begin{align}
\F{\mu\nu}\rho =\6_\mu\A\nu\rho-\6_\nu\A\mu\rho
\label{a3}
\end{align}
and Lorentz vector indices are lowered and raised by means of the Minkowski metric $\eta_{\mu\nu}=\mathrm{diag}(1,-1,-1,-1)$ and its inverse.
It can be shown that, up to total divergences, \eqref{a2} is in four spacetime dimensions the most general free field Lagrangian for the fields $\A\mu\nu$ which is quadratic in derivatives of these fields and invariant up to total divergences under the gauge transformations \eqref{a1} and standard global Poincar\'e transformations.

For later purpose I remark that \eqref{a2} also can be written as
\begin{align}
\LB=
a^\prime_1 F_{\mu\nu\rho}F^{\mu\nu\rho}
+a^\prime_2H_\mu H^\mu
+a_3G_\mu G^\mu
+a^\prime_4G_\mu H^\mu
\label{a4}
\end{align}
where 
\begin{align}
H^\mu=\tfrac 14\epsilon^{\mu\nu\rho\sigma}F_{\nu\rho\sigma},\quad 
G_\mu=\F{\mu\nu}\nu,\quad
a^\prime_1 =a_1-\tfrac 12 a_2,\quad
a^\prime_2=-4a_2,\quad
a^\prime_4=-4a_4
\label{a5}
\end{align}
There are particular values of the coefficients $a_1,\dots,a_4$ which deserve special attention. These values can be obtained from the equations of motion arising from $\LB$ by its Euler-Lagrange derivatives w.r.t. $\A\mu\nu$. To see this it is useful to decompose $A_{\mu\nu}$ into its symmetric and antisymmetric parts according to
\begin{align}
A_{\mu\nu}=H_{\mu\nu}+B_{\mu\nu},\quad
H_{\mu\nu}=H_{\nu\mu},\quad
B_{\mu\nu}=-B_{\nu\mu}
\label{a6}
\end{align}
The Euler-Lagrange derivatives of $\LB$ w.r.t. $H_{\mu\nu}$ and $B_{\mu\nu}$ are
\begin{align}
\EL{\LB}{H_{\mu\nu}}
=\,&
2(a_2-2a_1)\6_\rho\6^\rho H^{\mu\nu}
+2(2a_1-a_2-a_3)\6_\rho\6^{(\mu}H^{\nu)\rho}\nonumber\\
&+2a_3(\6^\mu\6^\nu H_\rho{}^\rho
+\eta^{\mu\nu}\6_\rho\6_\sigma H^{\rho\sigma}
-\eta^{\mu\nu}\6_\rho\6^\rho H_\sigma{}^\sigma)\nonumber\\
&-2(2a_1-a_2+a_3)\6_\rho\6^{(\mu}B^{\nu)\rho}
+2a_4\epsilon^{\rho\sigma\tau(\mu}\6^{\nu)}\6_\rho B_{\sigma\tau}
\label{a7}\\
\EL{\LB}{B_{\mu\nu}}
=\,&
2(2a_1-a_2+a_3)\6_\rho\6^{[\mu}H^{\nu]\rho}
+2a_4\epsilon^{\mu\nu\rho\sigma}\6^{\tau}\6_\rho H_{\sigma\tau}
\nonumber\\
&-2(2a_1+a_2)\6_\rho\6^\rho B^{\mu\nu}
-2(2a_1+3a_2-a_3)\6_\rho\6^{[\mu}B^{\nu]\rho}
\nonumber\\
&+2a_4(\epsilon^{\rho\sigma\tau[\mu}\6^{\nu]}+\epsilon^{\mu\nu\rho\sigma}\6^{\tau})\6_\rho B_{\sigma\tau}
\label{a8}
\end{align}
One observes that the equations of motion for $H_{\mu\nu}$ and $B_{\mu\nu}$ decouple if and only if $2a_1-a_2+a_3$ and $a_4$ vanish:
\begin{align}
a_3=&\,a_2-2a_1,\ a_4=0\ \Leftrightarrow\ a_3=-2a^\prime_1,\ a^\prime_4=0: \label{a9}\\
\EL{\LB}{H_{\mu\nu}}
=&\,
2a_3(\6_\rho\6^\rho H^{\mu\nu}
-2\6_\rho\6^{(\mu}H^{\nu)\rho}
+\6^\mu\6^\nu H_\rho{}^\rho
+\eta^{\mu\nu}\6_\rho\6_\sigma H^{\rho\sigma}
-\eta^{\mu\nu}\6_\rho\6^\rho H_\sigma{}^\sigma)\nonumber\\
\label{a10}\\
\EL{\LB}{B_{\mu\nu}}
=&\,
-2(2a_1+a_2)(\6_\rho\6^\rho B^{\mu\nu}+2\6_\rho\6^{[\mu}B^{\nu]\rho})
=-6(2a_1+a_2)\6_\rho \6^{[\rho} B^{\mu\nu]}
\label{a11}
\end{align}
The Euler-Lagrange derivatives \eqref{a10} are proportional to the left hand sides of the linearized vacuum Einstein equations in flat background. The Euler-Lagrange derivatives \eqref{a11} are proportional to the left hand sides of the standard Maxwell type free field equations for an antisymmetric gauge field $B_{\mu\nu}$. The cases \eqref{a9} are special because in these cases the free theory has two independent gauge symmetries, namely $\De0v H_{\mu\nu}=\6_{(\mu}v_{\nu)}$ and
$\De0w B_{\mu\nu}=\6_{[\mu}w_{\nu]}$ with unrelated gauge parameters $v_\mu$ and $w_\mu$ whereas for generic values of $a_1,\dots,a_4$ the gauge transformations of $H_{\mu\nu}$ and $B_{\mu\nu}$ in the free theory are $\De0v H_{\mu\nu}=\6_{(\mu}v_{\nu)}$ and
$\De0v B_{\mu\nu}=\6_{[\mu}v_{\nu]}$ with the same gauge parameters $v_\mu$.

The cases \eqref{a9} are even more special when $a_2=-2a_1$ or $a_2=2a_1$. In the first case the Euler-Lagrange derivatives of $\LB$ w.r.t. $B_{\mu\nu}$ vanish identically, i.e. up to a total divergence $\LB$  does not depend on $B_{\mu\nu}$ and thus effectively $B_{\mu\nu}$ drops out of the free theory.\footnote{This can be formulated as a gauge invariance of the free theory under arbitrary shifts of $B_{\mu\nu}$ which corresponds to the invariance of the Lagrangian \eqref{a24} under local Lorentz transformations for coefficients as in \eqref{a12}.} In this case the free theory is equivalent to linearized Einstein gravity in flat background. In the second case the Euler-Lagrange derivatives of $\LB$ w.r.t. $H_{\mu\nu}$ vanish identically, i.e. $H_{\mu\nu}$ drops out of the free theory. In that case the free theory is just a free theory for an antisymmetric gauge field $B_{\mu\nu}$ with Maxwell type Lagrangian proportional to $F_{\mu\nu\rho}F^{\mu\nu\rho}$ with $F_{\mu\nu\rho}=3\6_{[\mu} B_{\nu\rho]}$,
\begin{align}
&a_2=-2a_1,\ a_3=-4a_1,\ a_4=0\ \Leftrightarrow\ a^\prime_2=4a^\prime_1,\ a_3=-2a^\prime_1,\ a^\prime_4=0: \
\EL{\LB}{B_{\mu\nu}}=0
\label{a12}\\
&a_2=2a_1,\ a_3=a_4=0\ \Leftrightarrow\ a^\prime_1=a_3=a^\prime_4=0: \
\EL{\LB}{H_{\mu\nu}}=0
\label{a13}
\end{align}
So far the gauge fields $\A\mu\nu$ are not related to spacetime translations. In order to relate them to spacetime translations we use Noether couplings of these  gauge fields to the Noether currents corresponding via Noether's first theorem \cite{Noether:1918zz} to the global symmetries of the free field theory under ``improved'' spacetime translations. The generator of the improved spacetime translation in the $\mu$th spacetime direction is denoted by $\Delta_\mu$ and acts on the gauge fields according to
\begin{align}
\Delta_\mu\A\nu\rho=\F{\mu\nu}\rho
\label{a14}
\end{align}
$\Delta_\mu\A\nu\rho$ is the sum of a standard infinitesimal global spacetime translation of $\A\nu\rho$ in the $\mu$th spacetime direction generated by $\6_\mu\A\nu\rho$ and a gauge transformation \eqref{a1} with field dependent gauge parameters $-\A\mu\rho$. As both the standard global spacetime translations and the gauge transformations \eqref{a1} are symmetries of the free field theory, this also holds for the improved spacetime translations
generated by \eqref{a14}. The advantage of using the  improved spacetime translations in place of the usual spacetime translations here is that the corresponding Noether currents are invariant under the gauge transformations \eqref{a1}. The components $\nu$ of the Noether current corresponding to the global symmetry of the free theory generated by $\Delta_\mu$ are the components $T_\mu{}^\nu$ of the ``improved'' (gauge invariant) energy-momentum tensor
\begin{align}
T_\mu{}^\nu=&\,
\delta^\nu_\mu \LB-(\Delta_\mu\A\rho\sigma)\, \frac{\6 \LB}{\6(\6_\nu \A\rho\sigma)}\nonumber\\
=\,&
\delta^\nu_\mu \LB-\F{\mu\rho}\sigma (4a_1F^{\nu\rho}{}_\sigma+4a_2\F\sigma{[\nu\rho]}
+4a_3G^{[\nu}\delta^{\rho]}_\sigma
+2a_4\epsilon^{\nu\rho}{}_{\sigma\tau}G^\tau
-8a_4H^{[\nu}\delta^{\rho]}_\sigma)
\nonumber\\
=\,&
\delta^\nu_\mu[(a_1 -\tfrac 12 a_2)F_{\rho\sigma\tau}F^{\rho\sigma\tau}
+4a_2H_\rho H^\rho
+a_3G_\rho G^\rho]
\nonumber\\
&-4(a_1 -\tfrac 12 a_2)F_{\mu\rho\sigma}F^{\nu\rho\sigma}
+2a_2(F_{\rho\sigma\mu}\epsilon^{\nu\rho\sigma\tau}H_\tau-4H_\mu H^\nu)
\nonumber\\
&+2a_3(\F{\mu\rho}\nu G^\rho-G_\mu G^\nu)
+a_4(F_{\rho\sigma\mu}\epsilon^{\nu\rho\sigma\tau}G_\tau
-4\F{\mu\rho}\nu H^\rho)
\label{a15}
\end{align}
The Noether couplings of the gauge fields $\A\mu\nu$ to the improved energy-momentum tensor add interaction terms to the free field Lagrangian $\LB$ that provide a first order deformation $\LBo$ of this Lagrangian which is
\begin{align}
\LBo=\A\nu\mu T_\mu{}^\nu
\label{a16}
\end{align}
This deformation of the free field Lagrangian $\LB$
is accompanied by a corresponding deformation $\4\delta^{(1)}_v$ of the gauge transformations \eqref{a1} which is 
\begin{align}
\4\delta^{(1)}_v \A\mu\nu=v^\rho\Delta_\rho\A\mu\nu=v^\rho\F{\rho\mu}\nu
\label{a16a}
\end{align}
As $T_\mu{}^\nu$ is invariant under the gauge transformations \eqref{a1}, one has
\begin{align}
\4\delta^{(1)}_v \LB+\De0v \LBo\simeq 0
\label{a17}
\end{align}
where $\simeq$ denotes equality up to a total divergence. According to \eqref{a17},
$\LBo$ and $\4\delta^{(1)}_v$ provide a consistent first order deformation of the free theory, i.e. deformations 
$\LB+g\LBo$ and $\De0v+g\4\delta^{(1)}_v$ of the free field Lagrangian and gauge transformations with (constant but otherwise arbitrary) real deformation parameter $g$ such that the deformed Lagrangian is invariant under the deformed gauge transformations, up to a total divergence and up to terms that are quadratic in $g$. The relation of the deformed gauge transformations to general coordinate transformations is obtained by writing \eqref{a16a} as follows:
\begin{align}
\4\delta^{(1)}_v \A\mu\nu=v^\rho(\6_\rho\A\mu\nu-\6_\mu\A\rho\nu)=
v^\rho\6_\rho\A\mu\nu+(\6_\mu v^\rho)\A\rho\nu+\6_\mu(-v^\rho\A\rho\nu)
\label{a18}
\end{align}
\eqref{a18} shows that $\4\delta^{(1)}_v \A\mu\nu$ is the sum of a standard infinitesimal general coordinate transformation with parameters $v^\mu$ which ignores the second (upper) index of $\A\mu\nu$, and a gauge transformation \eqref{a1} with gauge parameters $-v^\rho\A\rho\nu$. The latter portion, i.e. $\6_\mu(-v^\rho\A\rho\nu)$, can be removed from $\4\delta^{(1)}_v \A\mu\nu$ because it does not contribute to $\4\delta^{(1)}_v \LB$ in \eqref{a17} as
$\LB$ is invariant under the gauge transformation \eqref{a1} (for arbitrary $v$'s). Hence, in place of $\4\delta^{(1)}_v \A\mu\nu=v^\rho\F{\rho\mu}\nu$ one may use 
\begin{align}
\De1v\A\mu\nu=v^\rho\6_\rho\A\mu\nu+(\6_\mu v^\rho)\A\rho\nu
\label{a19}
\end{align}
without amending $\LBo$, i.e. one has
\begin{align}
\De1v \LB+\De0v \LBo\simeq 0
\label{a20}
\end{align}
Now, $\De1v\A\mu\nu$ has the form of an infinitesimal general coordinate transformation of $\A\mu\nu$ which ignores the upper index $\nu$. However, $(\De0v+g\De1v)\A\mu\nu$ additionally contains the portion $\De0v\A\mu\nu=\6_\mu v^\nu$. This suggests to introduce a field with components $\delta^\nu_\mu+g\A\mu\nu$ because the deformed gauge transformation $\De0v+g\De1v$ of this field precisely has the form of a standard infinitesimal general coordinate transformation (with parameters $gv^\mu$) which ignores the upper index $\nu$. $\delta^\nu_\mu+g\A\mu\nu$ thus transforms under the deformed gauge transformations precisely like a tetrad field under infinitesimal general coordinate transformations. 

It is now obvious how one can extend the first order deformation of the free field theory to all orders. Namely, switching to standard notation, one can introduce a tetrad field $\viel a\mu$ whose lower index is treated as  a covariant world index of general coordinate transformations and whose upper index is treated as a contravariant Lorentz vector index (with generally globally realized Lorentz transformations). Then the torsion
\begin{align}
\T {\mu\nu}a=\6_\mu\viel a\nu-\6_\nu\viel a\mu
\label{a22}
\end{align}
is an antisymmetric tensor under general coordinate transformations with respect to its world indices $\mu,\nu$. 
Presuming that the tetrad is invertible one introduces also the inverse tetrad $\Viel a\mu$ which fulfills  
\begin{align}
\Viel a\mu\viel b\mu=\delta^b_a\,,\quad \Viel a\nu\viel a\mu=\delta^\nu_\mu
\label{a21}
\end{align}
Now one can convert the world indices of $\T {\mu\nu}a$ in the standard way into Lorentz vector indices according to
\begin{align}
\T {bc}a=\Viel b\mu\Viel c\nu(\6_\mu\viel a\nu-\6_\nu\viel a\mu)
\label{a23}
\end{align}
By construction $\T {bc}a$ transforms scalarly under general coordinate transformations. 
Hence the Lagrangian
\begin{align}
L_\mathrm{Bose}=\,&
e(a_1 T_{abc}T^{abc}
+a_2T_{abc}T^{cab}
+a_3\T{ab}b T^{ac}{}_c
+a_4\epsilon^{abcd}T_{abc}\T{de}e)
\nonumber\\
=\,&e(a^\prime_1 T_{abc}T^{abc}
+a^\prime_2\ch_a \ch^a
+a_3\cg_a \cg^a
+a^\prime_4\ch_a \cg^a),
\label{a24}
\end{align}
where $e=\det(\viel a\mu)$ denotes the determinant of the tetrad, Lorentz vector indices are lowered and raised by the Minkowski metric $\eta_{ab}$ and its inverse, $\ch^a=\tfrac 14\epsilon^{abcd}T_{bcd}$ and $\cg_a=\T{ab}b$, is a scalar density (with weight one) under general coordinate transformations and thus fulfills
\begin{align}
\delta_v L_\mathrm{Bose}=g\6_\mu (v^\mu L_\mathrm{Bose})
\label{a25}
\end{align}
where $\delta_v$ is an infinitesimal general coordinate transformation with parameters $gv^\mu$,
\begin{align}
\delta_v \viel a\mu=gv^\nu\6_\nu \viel a\mu+g(\6_\mu v^\nu) \viel a\nu
\label{a26}
\end{align}
Using $\viel a\mu=\delta^a_\mu+g\A\mu{a}$, an expansion of $g^{-2}L_\mathrm{Bose}$ in powers of $g$ reproduces at zeroth order in $g$ the free field Lagrangian \eqref{a2} and at first order in $g$ the first order deformation \eqref{a16} of that Lagrangian. Furthermore, \eqref{a26} provides \eqref{a1} and \eqref{a19} via $\viel a\mu=\delta^a_\mu+g\A\mu{a}$. Hence \eqref{a24} (times $g^{-2}$) and \eqref{a26} complete the first order deformations \eqref{a16} and \eqref{a19} of the free field Lagrangian \eqref{a2} and gauge transformations \eqref{a1} to all orders in $g$ via $\viel a\mu=\delta^a_\mu+g\A\mu{a}$. Of course, the expansion of $g^{-2}L_\mathrm{Bose}$ in $g$ yields infinitely many terms owing to the presence of the inverse tetrad in $L_\mathrm{Bose}$. As is well-known, possibly first observed by Lanczos (cf. section  5.1 of \cite{Sauer:2004hj}), the Lagrangian \eqref{a24} is for $a_2=-2a_1$, $a_3=-4a_1$, $a_4=0$ (the so-called ``teleparallel equivalent of general relativity'') proportional to the Einstein-Hilbert Lagrangian for the metric $g_{\mu\nu}=\viel a\mu\viel b\nu\eta_{ab}$, up to a total divergence.\footnote{A corresponding ``teleparallel equivalent of supergravity'' is given in section 1.5 of \cite{VanNieuwenhuizen:1981ae}.\label{sugra}}

\mysection{Supersymmetric free field theories}\label{susyfree}

In order to construct supersymmetric theories of gauged spacetime translations, we first seek supersymmetric extensions of free field theories with a Lagrangian \eqref{a2} in flat spacetime. To this end we introduce fermionic spinor-vector fields $\lambda_\alpha{}^\nu$ which are to become the superpartner fields of the gauge fields $\A\mu\nu$. The index $\alpha$ of $\lambda_\alpha{}^\nu$ is a spinor index with two values ($\alpha=1,2$, cf. appendix \ref{conv}), the index $\nu$ is a contravariant Lorentz vector index (in flat spacetime). 
$\lambda_\alpha{}^\nu$ is a complex-valued field (i.e. a field with a real part and an imaginary part) which is invariant under the gauge transformations \eqref{a1},
\begin{align}
\De 0v\lambda_\alpha{}^\nu=0
\label{b0}
\end{align}
The complex conjugate of $\lambda_\alpha{}^\nu$ is denoted by $\5\lambda_\da{}^\nu$ ($\da=\dot{1},\dot{2}$).

We denote the generators of global supersymmetry transformations in flat spacetime by $D_\alpha$ and $\5 D_\da$ where $\5 D_\da$ is the complex conjugate of $D_\alpha$, and make the following Ans\"atze for the supersymmetry transformations $D_\alpha$ of $A_{\mu\nu}$, $\lambda^\beta{}_\mu$ and $\5\lambda^\da{}_\mu$:
\begin{align}
D_\alpha A_{\mu\nu}=\,&
(b_1\sigma_\mu\5\lambda_\nu+b_2\sigma_\nu\5\lambda_\mu
+b_3\eta_{\mu\nu}\sigma^\rho\5\lambda_\rho
+b_4\epsilon_{\mu\nu\rho\sigma}\sigma^\rho\5\lambda^\sigma)_\alpha
\label{b1}\\
D_\alpha \lambda^\beta{}_\mu=\,&z[\delta_\alpha^\beta (b_5 G_\mu+b_6 H_\mu)
+\sigma_{\mu\nu\alpha}{}^\beta (b_7 G^\nu+b_8 H^\nu)
+b_9\sigma^{\nu\rho}{}_\alpha{}^\beta F_{\nu\rho\mu}]
\label{b2}\\
D_\alpha\5\lambda^\da{}_\mu=\,&0
\label{b3}
\end{align}
where the coefficients $b_1,\dots,b_9$ in general are complex numbers, i.e.
\[
b_i=x_i+\Ii y_i,\quad x_i,y_i\in\mathbb{R}\quad (i=1,\ldots,9)
\]
The Ans\"atze \eqref{b1} through \eqref{b3} are motivated by the requirements that the supersymmetry transformations $D_\alpha$ of $A_{\mu\nu}$, $\lambda^\beta{}_\mu$ and $\5\lambda^\da{}_\mu$ do not explicitly depend on spacetime coordinates and commute with the gauge transformations \eqref{a1}.

For the supersymmetric free field Lagrangrian we make the following Ansatz:
\begin{align}
\LS=\,&\LB+\LF
\label{b5}\\
\LF=\,&
z[a_5\lambda^\mu\sigma^\nu\6_\nu\5\lambda_\mu
+a_6(\lambda^\mu\sigma^\nu\6_\mu\5\lambda_\nu+\lambda^\mu\sigma_\mu\6_\nu\5\lambda^\nu)
\nonumber\\
&+\Ii a_7(\lambda^\mu\sigma^\nu\6_\mu\5\lambda_\nu-\lambda^\mu\sigma_\mu\6_\nu\5\lambda^\nu)
+\Ii a_8\epsilon^{\mu\nu\rho\sigma}\lambda_\mu\sigma_\nu\6_\rho\5\lambda_\sigma]
\label{b6}
\end{align}
with $\LB$ as in \eqref{a2} and real coefficients $a_5,\dots,a_8$. $\LF$ is the most general free field Lagrangian for $\lambda^\alpha{}_\mu$ and $\5\lambda^\da{}_\mu$ that is invariant under standard Poincar\'e transformations, linear in the derivatives of these fields and real, up to total divergences. 

We impose that the anticommutators of the $D_\alpha$ and $\5D_\da$ fulfill the standard supersymmetry algebra, i.e.
\begin{align}
\acom {D_\alpha}{\5D_\da}\sim -2z\sigma^\mu{}_{\alpha\da}\6_\mu\,,\quad
\acom {D_\alpha}{D_\beta}\sim 0,\quad
\acom {\5D_\da}{\5D_\dbe}\sim 0
\label{b4}
\end{align}
where $\sim$ denotes equality up to gauge transformations \eqref{a1} (with field dependent parameters) and up to terms that vanish on-shell in the free field theory, i.e. up to terms containing the Euler-Lagrange derivatives of $\LS$ w.r.t. $\A\mu\nu$, $\lambda_\alpha{}^\nu$ or 
$\5\lambda_\da{}^\nu$. \eqref{b5} implies that the Euler-Lagrange derivatives of $\LS$ w.r.t. $\A\mu\nu$ are of second order in the derivatives of $\A\mu\nu$. Hence, the Ans\"atze \eqref{b1} through \eqref{b3} imply that the anticommutators $\acom {D_\alpha}{\5D_\da}A_{\mu\nu}$ do not contain the Euler-Lagrange derivatives of $\LS$ w.r.t. $\A\mu\nu$ when we require that these anticommutators do not contain gauge transformations or on-shell vanishing terms that explicitly depend on spacetime coordinates.\footnote{This requirement appears to be natural but it is not completely innocent as there are supersymmetric free field theories where the commutator algebra of the infinitesimal supersymmetry transformations contains infinitesimal gauge transformations and on-shell vanishing terms which depend on spacetime coordinates explicitly even though the supersymmetry transformations themselves do not depend on spacetime coordinates explicitly, cf. \cite{Brandt:2000uw,Brandt:2000ky} for examples.}
With this requirement, \eqref{b4} imposes on $A_{\mu\nu}$ for generic values of $a_1,\dots,a_4$, i.e. for values which exclude the special cases \eqref{a9}:\footnote{In the cases \eqref{a9} one has $\acom {D_\alpha}{\5D_\da}H_{\mu\nu}=-2z\sigma^\rho{}_{\alpha\da}\6_\rho H_{\mu\nu}+\6_{(\mu} X_{\nu)\alpha\da}$ and $\acom {D_\alpha}{\5D_\da}B_{\mu\nu}=-2z\sigma^\rho{}_{\alpha\da}\6_\rho B_{\mu\nu}+\6_{[\mu} Y_{\nu]\alpha\da}$ where $X_{\mu\alpha\da}$ and $Y_{\mu\alpha\da}$ can be different. In the very special case  \eqref{a12}, which provides linearized standard supergravity, $\acom {D_\alpha}{\5D_\da}B_{\mu\nu}$ does not impose any condition at all because then $B_{\mu\nu}$ effectively drops out of the free field theory, cf. remarks following \eqref{a11}. For these reasons, the coefficients $b_i$ which occur in linearized standard supergravity need not and actually do not fulfill all of the equations in appendix \ref{algebra}. This is the reason why, for instance, the coefficients in \eqref{b34} do not provide linearized standard supergravity.}
\begin{align}
\acom {D_\alpha}{\5D_\da}A_{\mu\nu}= -2z\sigma^\rho{}_{\alpha\da}\6_\rho A_{\mu\nu}+\6_\mu X_{\nu\alpha\da}
\label{b7}
\end{align}
where $ X_{\nu\alpha\da}$ are some field dependent gauge parameters of a gauge transformation \eqref{a1}. In appendix \ref{algebra} it is shown that \eqref{b7} and the Ans\"atze \eqref{b1} through \eqref{b3} imply that with no loss of generality one may presume $b_9=-1$ and then obtains
\begin{align}
X_{\nu\alpha\da}=2z A_{\rho\nu}\sigma^\rho{}_{\alpha\da}
\label{b11}
\end{align}
Furthermore in appendix \ref{algebra} it is shown that the choice $y_1=y_3=0$ provides the following coefficients which will be used in the further investigations:
\begin{align}
&b_1\in\mathbb{R}\backslash \{\tfrac 34\},\ y_5,y_6,y_7,y_8\in\mathbb{R},
\nonumber\\
&b_2=1-b_1,\ b_3=b_1-1,\ b_4=\Ii (1-b_1),
\nonumber\\
&b_5=\frac{1-b_1}{3-4b_1}+\Ii y_5,\ 
b_6=\Ii y_6,\ 
b_7=\frac{2(b_1-1)}{3-4b_1}+\Ii y_7,\ 
b_8=\Ii y_8,\
b_9=-1,
\nonumber\\
&(1-b_1)y_5=(1-\tfrac 32 b_1)y_7,\ 
(1-b_1)(2+y_6)=(1-\tfrac 32 b_1)y_8
\label{b34}
\end{align}
The values of $y_5,y_6,y_7,y_8$ in \eqref{b34} are so far only restricted by the last two equations in \eqref{b34} -- hence, for any particular value of $b_1$, at most two of the coefficients $y_5,y_6,y_7,y_8$ are independent.  Furthermore the value $b_1=3/4$ must be excluded in the cases $y_1=0$ (cf. derivation of \eqref{b28} in appendix \ref{algebra}).

Of course, in addition to \eqref{b7}, the algebra \eqref{b4} is required to be satisfied on $\lambda_\alpha{}^\nu$ and $\5\lambda_\da{}^\nu$. However, on $\lambda_\alpha{}^\nu$ and $\5\lambda_\da{}^\nu$, the anticommutators of the supersymmetry transformations can (and do) involve the Euler-Lagrange derivatives of $\LS$ w.r.t. these fields. Therefore it appears to be more efficient to first determine Lagrangians $\LS$  which are invariant, up to a total divergence, under the supersymmetry transformations with coefficients fulfilling \eqref{b34}. The algebra \eqref{b4} then ``automatically'' will also hold on $\lambda_\alpha{}^\nu$ and $\5\lambda_\da{}^\nu$.

Now, as $\LS$ is real up to a total divergence, it is annihilated up to a total divergence by $\5D_\da$ whenever it is annihilated up to a total divergence by $D_\alpha$. Hence, it is sufficient to consider $D_\alpha \LS$. Using \eqref{b1} through \eqref{b3}, one obtains 
\begin{align}
D_\alpha \LS\simeq\,&[c_1\sigma_\nu\5\lambda_\mu\6_\rho F^{\rho(\mu\nu)}
+c_2\sigma_\mu \5\lambda^\mu \6_\nu G^\nu
+c_3\sigma^\nu \5\lambda^\mu\6_\nu H_\mu
+c_4\sigma^\nu \5\lambda^\mu \6_\mu H_\nu
\nonumber\\
&+c_5\sigma^\nu \5\lambda^\mu\6_\nu G_\mu
+c_6\sigma^\nu \5\lambda^\mu \6_\mu G_\nu
+\epsilon^{\mu\nu\rho\sigma}\sigma_\sigma\5\lambda_\mu(c_7\6_\nu H_\rho+c_8\6_\nu G_\rho)]_\alpha
\label{c1}
\end{align}
with (for arbitrary coefficients $b_i$)
\begin{align}
c_1=\,&z^2  (a_5+a_8)b_9-4a^\prime_1(b_1+b_2)
\nonumber\\
c_2=\,&-z^2[(a_6+\Ii a_7)b_5+(\tfrac 12 a_5-a_8)b_7+ a_8b_9]-4a^\prime_1 b_3-2a_3(b_1+b_2+3b_3)
\nonumber\\
c_3=\,&-z^2[(\tfrac 12 a_6+\tfrac {\Ii}2 a_7+a_8)b_8+a_5 b_6+2\Ii a_8 b_9]
+8a^\prime_1 b_4-2a^\prime_2b_4+a^\prime_4 b_2
\nonumber\\
c_4=\,&z^2[(\tfrac 12 a_5+2a_6-\Ii a_7)b_8-(a_6-\Ii a_7)b_6 +2(\Ii a_6+a_7)b_9]
-8a^\prime_1 b_4+2a^\prime_2 b_4+a^\prime_4 b_1
\nonumber\\
c_5=\,&-z^2[(\tfrac 12 a_6+\tfrac {\Ii}2  a_7+a_8)b_7
+a_5 b_5-(\tfrac 12 a_5+\tfrac 12 a_8+a_6 +\Ii a_7)b_9]
\nonumber\\
&
+2a^\prime_1(b_2-b_1)-a^\prime_4 b_4+2a_3b_2
\nonumber\\
c_6=\,&z^2[(\tfrac 12 a_5+2a_6-\Ii a_7)b_7-(a_6-\Ii a_7)b_5-(\tfrac 12 a_5-\tfrac 12 a_8+2a_6)b_9]
\nonumber\\
&+2a^\prime_1(b_1-b_2)+a^\prime_4 b_4+2a_3b_1
\nonumber\\
c_7=\,&
z^2[\tfrac 12(\Ii a_5-\Ii a_6+a_7-\Ii a_8)b_8 +\Ii a_8 b_6 -(a_5-a_8)b_9]
\nonumber\\
&+4a^\prime_1(b_1-b_2)+a^\prime_2(b_2-b_1)+a^\prime_4 b_4
\nonumber\\
c_8=\,&
z^2[\tfrac 12(\Ii a_5-\Ii a_6+a_7-\Ii a_8)b_7 +\Ii a_8 b_5 +(\Ii a_6- a_7)b_9]
\nonumber\\
&+4a^\prime_1 b_4+\tfrac 12 a^\prime_4(b_2-b_1)+2a_3 b_4
\label{c2}
\end{align}
All the coefficients $c_1,\dots,c_8$ must vanish in the generic cases. We shall now sum up what this imposes in the cases of
coefficients $b_1,\dots,b_9$ given in \eqref{b34}. In these cases, obviously
$c_1=0$ is a real equation and provides \eqref{11.a1}:
\begin{align}
a^\prime_1=
-\frac{z^2(a_5+a_8)}4
\label{11.a1}
\end{align}
$c_7-\Ii c_3=0$ turns out to be a real equation too which, using \eqref{11.a1}, provides \eqref{11.a2}:
\begin{align}
a^\prime_2=
-z^2 (a_5+a_8)(y_6+\tfrac 12 y_8) 
\label{11.a2}
\end{align}
$c_8-\Ii c_5=0$ also turns out to be a real equation which, using \eqref{11.a1}, provides \eqref{11.a4}:
\begin{align}
a^\prime_4=
-2z^2 (a_5+a_8)(y_5+\tfrac 12 y_7) 
\label{11.a4}
\end{align}
$c_5+c_6=0$ is a complex equation which provides \eqref{11.a3} by its real part and \eqref{11.10} by its imaginary part:
\begin{align}
&a_3=
-\frac {z^2}2\left[ \frac {2(1-b_1)a_5+a_6+(1-2b_1)a_8}{4b_1-3}
+a_7(\tfrac 32 y_7-y_5) \right]
\label{11.a3}\\
&\frac{a_7}{3-4b_1}+a_5(\tfrac 12 y_7-y_5)+a_6(\tfrac 32 y_7-y_5)-a_8y_7=0
\label{11.10}
\end{align}
$c_7+\Ii c_3=0$ is a complex equation which, using \eqref{11.a1}, \eqref{11.a2} and \eqref{11.a4}, provides \eqref{11.3} by its real part and \eqref{11.2} by its imaginary part:
\begin{align}
&2a_5[(1-b_1)(2-y_8)+(2b_1-1)y_6]
+2a_8[2(b_1-1)y_6+b_1(y_8-2)]+a_6y_8=0
\label{11.3}\\
&a_7y_8=4(1-b_1)(a_5+a_8)(y_5+\tfrac 12 y_7) 
\label{11.2}
\end{align}
$c_2=0$ is a complex equation which, using \eqref{11.a1} and \eqref{11.a3}, provides \eqref{11.12} by its real part and \eqref{11.11} by its imaginary part:
\begin{align}
&-2(1-b_1)^2a_5+(2b_1-1)a_6+2b_1(1-b_1)a_8=0
\label{11.12}\\
&-a_6y_5+\frac{(1-b_1)a_7}{4b_1-3}-(\tfrac 12 a_5-a_8)y_7=0
\label{11.11}
\end{align}
$c_3+c_4=0$ is a complex equation which, using \eqref{11.a1}, \eqref{11.a2} and \eqref{11.a4}, provides \eqref{11.8} by its real part and \eqref{11.7} by its imaginary part:
\begin{align}
&a_7(\tfrac 32 y_8-y_6-2)-2(a_5+a_8)(y_5+\tfrac 12 y_7)=0
\label{11.8}\\
&a_5(\tfrac 12 y_8-y_6)+a_6(\tfrac 32 y_8-y_6-2)+a_8(2-y_8)=0
\label{11.7}
\end{align}
Finally, $c_8+\Ii c_5=0$ is a complex equation which, using \eqref{11.a4}, provides \eqref{11.6} by its real part.
 Using \eqref{11.a1} and \eqref{11.a3}, the imaginary part of  $c_8+\Ii c_5=0$ gives again \eqref{11.12}.
\begin{align}
\frac{2(3b_1-2)a_7}{4b_1-3}+2a_5[(2b_1-1)y_5+(b_1-1)y_7]+a_6y_7
+2a_8[2(b_1-1)y_5+b_1y_7]=0
\label{11.6}
\end{align}
We note that for $b_1\neq 1$ we have $\tfrac 32 y_8-y_6-2=\tfrac 12 y_8/(1-b_1)$ by the last equation in \eqref{b34} which then implies that \eqref{11.2} and \eqref{11.8} are equivalent. For $b_1=1$ one has $y_8=0$ by the last equation in \eqref{b34} and thus \eqref{11.2} becomes trivial in the case $b_1=1$. Hence, actually \eqref{11.2} does not provide any extra condition in either case and can be ignored.

Notice that \eqref{11.10} through \eqref{11.6} have been written such that they only involve the coefficients $a_5,\dots,a_8$ in $\LF$ but not those in $\LB$. Hence, \eqref{11.10} through \eqref{11.6} are equations for $a_5,\dots,a_8$ and $b_1,y_5,\dots,y_8$ which must be fulfilled in order that $\LS$ is invariant, up to a total divergence, under the supersymmetry transformations generated by \eqref{b1} through \eqref{b3} with coefficients $b_i$ as in \eqref{b34}. 
\eqref{11.a1} through \eqref{11.a3} then provide the coefficients $a^\prime_1$, $a^\prime_2$, $a_3$, $a^\prime_4$ 
in $\LB$, without imposing further conditions. In other words, in order to obtain a supersymmetric free field theory with a Lagrangian $\LS$ and supersymmetry transformations \eqref{b1} through \eqref{b3} with coefficients $b_i$ as in \eqref{b34}, it is necessary and sufficient that equations \eqref{11.10} through \eqref{11.6} (where one can ignore \eqref{11.2}) and additionally the last two equations in \eqref{b34} are fulfilled.

We shall now provide various supersymmetric free field theories arising from these equations by giving the respective coefficients $a^\prime_1,\dots,a_8$ in the Lagrangian $\LS$ and $b_1,\dots,b_9$ in the supersymmetry transformations \eqref{b1} through \eqref{b3}. Firstly we provide the free field theories for the cases $b_1=1$ because for $b_1=1$ both the supersymmetry transformations \eqref{b1} through \eqref{b3} (with coefficients $b_i$ as in \eqref{b34}) and the equations \eqref{11.10} through \eqref{11.6} are particularly simple, and the last two equations in \eqref{b34} impose $y_7=y_8=0$. One obtains:
\begin{align}
&a_5,y_5,y_6\in \mathbb{R},
\nonumber\\
&b_1=1,\ b_2=b_3=b_4=0,\ b_5=\Ii y_5,\ b_6=\Ii y_6,\ b_7=b_8=0,\ b_9=-1,
\nonumber\\
&a^\prime_1=-\tfrac 18 z^2a_5(2+y_6),\
a^\prime_2=-\tfrac 12 z^2a_5y_6(2+y_6),\
a_3=\tfrac 12 z^2a_5(-y_5^2+\tfrac 12 y_6),
\nonumber\\
&a^\prime_4=-z^2a_5y_5(2+y_6),\
a_6=0,\
a_7=-a_5y_5,\
a_8=\tfrac 12 a_5y_6 
\label{sol1}
\end{align}
Secondly we provide the free field theories for the cases $b_1=2/3$. 
In these cases one obtains:\footnote{The case $a_5=2a_8$ is excluded in \eqref{sol2} because $b_1=2/3$ and $a_5=2a_8$ imply that all coefficients $a_1,\dots,a_8$ must vanish.}
\begin{align}
&a_5,a_8,y_7\in \mathbb{R},\ a_5\neq 2a_8,
\nonumber\\
&b_1=\tfrac 23,\ 
b_2=\tfrac 13,\ 
b_3=-\tfrac 13,\
b_4=\tfrac{\Ii}3 ,\ 
b_5=1,\ 
b_6=-2\Ii ,\ 
b_7=-2+\Ii y_7,
\nonumber\\
&b_8=\frac{4\Ii(a_5+a_8)}{3(2a_8-a_5)},\ 
b_9=-1,\ 
a^\prime_1=-\tfrac 14 z^2(a_5+a_8),\
a^\prime_2=\frac{2z^2(a_5+a_8)(5a_8-4a_5)}{3(2a_8-a_5)},
\nonumber\\
&a_3=-\tfrac 12 z^2[5a_8-4a_5+\tfrac 34(2a_8-a_5)y_7^2],\
a^\prime_4=-z^2(a_5+a_8)y_7,
\nonumber\\
&
a_6=\tfrac 23(a_5-2a_8),\
a_7=\tfrac 12(2a_8-a_5)y_7
\label{sol2}
\end{align}
Thirdly we provide the free field theories for the cases $y_6=0$ with $b_1$ different from 1 and 2/3:
\begin{align}
&b_1\in \mathbb{R}\backslash\{1,\tfrac 23,\tfrac 34\},\ a_8,y_7\in \mathbb{R},
\nonumber\\
&
b_2=1-b_1,\ 
b_3=b_1-1,\
b_4=\Ii(1-b_1) ,\ 
b_5=\frac{1-b_1}{3-4b_1}+\frac{\Ii (2-3b_1)y_7}{2(1-b_1)},\ 
b_6=0,
\nonumber\\
&b_7=\frac{2(b_1-1)}{3-4b_1}+\Ii y_7,\ 
b_8=\frac{4\Ii (b_1-1)}{3b_1-2},\
b_9=-1,
\nonumber\\
&a^\prime_1=-\frac {z^2a_8}{4(1-b_1)},\
a^\prime_2=\frac{2z^2a_8}{3b_1-2},
\nonumber\\
&a_3=-z^2a_8\left[ \frac{1}{2(4b_1-3)}+\frac{(3b_1-2)(4b_1-3)y_7^2}{8(1-b_1)^3}\right],\
a^\prime_4=\frac{z^2(4b_1-3)a_8y_7}{(1-b_1)^2},
\nonumber\\
&
a_5=\frac{b_1a_8}{(1-b_1)},\ 
a_6=0,\
a_7=\frac {(3b_1-2)(4b_1-3)a_8y_7}{2(1-b_1)^2}
\label{sol3}
\end{align} 
\eqref{sol1} through \eqref{sol3} show that there are supersymmetric extensions of various free field theories with a Lagrangian $\LB$ as in \eqref{a2}, where of course supersymmetry relates the coefficients $a_1,\dots,a_4$ of $\LB$. Furthermore, we remark that \eqref{sol1} through \eqref{sol3} do not exhaust such free field theories because there are other free field theories with coefficients 
$a_1,\dots,a_8 $ and $b_1,\dots,b_9$ different from those in \eqref{sol1}, \eqref{sol2}, \eqref{sol3} which fulfill equations \eqref{b34} and \eqref{11.10} through \eqref{11.6} (we leave it to interested readers to work out such free field theories).

We end the discussion of supersymmetric free field theories by addressing 
the issue of auxiliary fields which might be used to close the commutator algebra of the infinitesimal supersymmetry transformations and spacetime translations off-shell, up to gauge transformations. According to the standard counting of degrees of freedom, the free field theory with Lagrangian $\LS$ has, in the generic cases, 12 bosonic and 16 fermionic degrees of freedom off-shell. Hence one needs at least four additional bosonic degrees of freedom to close the commutator algebra of the infinitesimal supersymmetry transformations and spacetime translations off-shell, up to gauge transformations. This suggests to introduce an auxiliary real vector field $B_\mu$ and the following Ansatz for modified supersymmetry transformations $D^\prime_\alpha$ in presence of $B_\mu$:
\begin{align}
D^\prime_\alpha A_{\mu\nu}=\,&D_\alpha A_{\mu\nu}=(b_1\sigma_\mu\5\lambda_\nu+b_2\sigma_\nu\5\lambda_\mu
+b_3\eta_{\mu\nu}\sigma^\rho\5\lambda_\rho
+b_4\epsilon_{\mu\nu\rho\sigma}\sigma^\rho\5\lambda^\sigma)_\alpha
\label{b1a}\\
D^\prime_\alpha \lambda^\beta{}_\mu=\,&z[\delta_\alpha^\beta (b_5 G_\mu+b_6 H_\mu+b_{10} B_\mu)
+\sigma_{\mu\nu\alpha}{}^\beta (b_7 G^\nu+b_8 H^\nu+b_{11}B^\nu)
+b_9\sigma^{\nu\rho}{}_\alpha{}^\beta F_{\nu\rho\mu}]
\label{b2a}\\
D^\prime_\alpha\5\lambda^\da{}_\mu=\,&D_\alpha\5\lambda^\da{}_\mu=0
\label{b3a}
\end{align}
where $b_{10}$ and $b_{11}$ are complex coefficients,
\begin{align}
b_{10}=x_{10}+\Ii y_{10},\ b_{11}=x_{11}+\Ii y_{11},\ x_{10},x_{11},y_{10},y_{11}\in\mathbb{R}
\label{b4a}
\end{align}
Now, as the algebra \eqref{b4} was already realized off-shell on $A_{\mu\nu}$, 
the $B_\mu$-dependent terms arising from \eqref{b2a} must not contribute in 
$\acom {D^\prime_\alpha}{\5D^\prime_\da}A_{\mu\nu}$. Up to a factor $z$ these terms turn out to be
\begin{align}
&B_\nu\sigma_{\mu\alpha\da}(b_1b_{10}+\tfrac 12 b_2b_{11}+\Ii \5b_4b_{11})
\nonumber\\
&+B_\mu\sigma_{\nu\alpha\da}(b_2b_{10}+\tfrac 12 b_1b_{11}-\Ii \5b_4b_{11})
\nonumber\\
&+\eta_{\mu\nu}B_\rho \sigma^\rho{}_{\alpha\da}[\5b_3(b_{10}-\tfrac 32 b_{11})
-\tfrac 12 (b_1+b_2)b_{11}]
\nonumber\\
&+\epsilon_{\mu\nu\rho\sigma}B^\rho\sigma^\sigma{}_{\alpha\da}
[\5b_4(-b_{10}+\tfrac 12 b_{11})
+\tfrac \Ii 2 (b_1-b_2)b_{11}]+c.c.
\label{b5a}
\end{align}
It is straightforward to verify that all terms in \eqref{b5a} vanish for
coefficients $b_i$ as in \eqref{b34} if and only if
\begin{align}
x_{10}=x_ {11}=0,\ (1-b_1)y_{10}=(1-\tfrac 32 b_1)y_{11}
\label{b6a}
\end{align}
\eqref{b6a} thus are necessary conditions for the commutator algebra of infinitesimal spacetime translations and supersymmetry transformations 
with $D^\prime_\alpha$ as in \eqref{b1a} through \eqref{b3a} to close off-shell, up to gauge transformations \eqref{a1}, with coefficients $b_i$ as in \eqref{b34}. 
In the case $b_1=1$ one has $y_{11}=0$ and obtains that the commutator algebra of the infinitesimal supersymmetry transformations and spacetime translations closes off-shell, up to gauge transformations \eqref{a1}, for the following transformations $D^\prime_\alpha$:
\begin{align}
b_1=1:\ 
&D^\prime_\alpha A_{\mu\nu}=(\sigma_\mu\5\lambda_\nu)_\alpha
\label{b1aa}\\
&D^\prime_\alpha \lambda^\beta{}_\mu=z(\Ii\delta_\alpha^\beta  \7B_\mu
-\sigma^{\nu\rho}{}_\alpha{}^\beta F_{\nu\rho\mu})
\label{b2aa}\\
&D^\prime_\alpha\5\lambda^\da{}_\mu=0
\label{b3aa}\\
&D^\prime_\alpha \7B_\mu=-\Ii (\sigma^\nu \6_\nu\5\lambda_\mu)_\alpha
\label{b4aa}
\end{align}
where, in order to simplify the supersymmetry transformations, we have redefined the auxiliary field according to
\begin{align}
\7B_\mu=y_5G_\mu+y_6 H_\mu+y_{10} B_\mu
\label{b7a}
\end{align}
In the cases $b_1\neq 1$ we write the transformations $D^\prime_\alpha$ in \eqref{b1a} through \eqref{b3a}, for coefficients $b_i$ as in \eqref{b34}, as follows:
\begin{align}
b_1\neq 1:\ 
D^\prime_\alpha A_{\mu\nu}=\,&
b_1(\sigma_\mu\5\lambda_\nu)_\alpha+(1-b_1)(\sigma_\nu\5\lambda_\mu
-\eta_{\mu\nu}\sigma^\rho\5\lambda_\rho
+\Ii\epsilon_{\mu\nu\rho\sigma}\sigma^\rho\5\lambda^\sigma)_\alpha
\label{b1aaaa}\\
D^\prime_\alpha \lambda^\beta{}_\mu=\,&
z\left[\delta_\alpha^\beta \left(\frac{1-b_1}{3-4b_1}G_\mu-2\Ii H_\mu+\Ii\,\frac{2-3b_1}{2(1-b_1)}\7B_\mu\right)\right.
\nonumber\\
&
\left.+\,\sigma_{\mu\nu\alpha}{}^\beta \left(-2\,\frac{1-b_1}{3-4b_1}G^\nu+\Ii \7B^{\nu}\right)
-\sigma^{\nu\rho}{}_\alpha{}^\beta F_{\nu\rho\mu}\right]
\label{b2aaaa}\\
D^\prime_\alpha\5\lambda^\da{}_\mu=\,&0
\label{b3aaaa}
\end{align}
where we have redefined the auxiliary field according to
\begin{align}
\7B_\mu=y_7G_\mu+y_8H_\mu +y_{11} B_\mu
\label{b8a}
\end{align}
Furthermore we make the following Ansatz for $D^\prime_\alpha \7B_\mu$:
\begin{align}
b_1\neq 1:\ 
D^\prime_\alpha\7B_\mu=\Ii(b_{12}\sigma^\nu\6_\mu\5\lambda_\nu
+b_{13}\sigma_\mu\6_\nu\5\lambda^\nu
+b_{14}\sigma^\nu\6_\nu\5\lambda_\mu)_\alpha
+b_{15}\epsilon_{\mu\nu\rho\sigma}(\sigma^\nu\6^\rho\5\lambda^\sigma)_\alpha
\label{b4aaaa}
\end{align}
where $b_{12},\dots,b_{15}$ are complex coefficients. Imposing now that $D^\prime_\alpha$ and $\5D^\prime_\da$
fulfill $\acom{D^\prime_\alpha}{\5D^\prime_\da}=-2z\sigma^\nu{}_{\alpha\da}\6_\nu$ and 
$\acom{D^\prime_\alpha}{D^\prime_\beta}=0$
on $\lambda^\gamma{}_\mu$ and $\7B_\mu$, one gets
\begin{align}
b_{12}=-\frac{2(1-b_1)^2}{4b_1-3},\qd
b_{13}=\frac{6(1-b_1)^2}{4b_1-3},\qd
b_{14}=b_{15}=\frac{2(1-b_1)(3b_1-2)}{4b_1-3}
\label{b8c}
\end{align}
I note that this result for $b_1\neq 1$ can be obtained from the result for $b_1=1$ presented in equations \eqref{b1aa} through \eqref{b4aa} by the following redefinitions of $\lambda_\mu$ and $\7B_\mu$ which comprise a redefinition of $\lambda_\mu$ as used in section  \ref{add} and a corresponding redefinition of $\7B_\mu$ such that \eqref{b2aa} yields \eqref{b2aaaa} for the primed fields:
\[
\lambda^\prime_\mu=\frac 1{4b_1-3}[2(1-b_1)\lambda^\nu\sigma_{\nu\mu}+(3b_1-2)\lambda_\mu],\qd
\7B^\prime_\mu=\frac{b_1-1}{4b_1-3}(2\7B_\mu+4H_\mu)
\]
Hence, for coefficients $b_i$ as in \eqref{b34}, the commutator algebra of infinitesimal supersymmetry transformations and spacetime translations can be closed off-shell, up to gauge transformations, by means of an auxiliary vector field for all values of $b_1$.\footnote{This corrects a statement in a previous version of this paper that the commutator algebra can be closed off-shell in this way only for $b_1=1$ and $b_1=2/3$.} 



\mysection{Supersymmetric actions for interacting fields}\label{Lnl}

\subsection{Lagrangians for ``teleparallel'' theories}\label{tele}

We now discuss the construction of globally supersymmetric extensions of actions with Lagrangians \eqref{a24} and generalizations thereof (containing higher powers of the torsion and/or terms with more derivatives). We denote the generators of the global supersymmetry transformations by $\cD_\alpha$ and $\5\cD_\da$ and define covariant derivatives $\cD_a$ according to
\begin{align}
\cD_a=\Viel a\mu\6_\mu
\label{d1}
\end{align}
We shall now presume that $\cD_\alpha$ and $\5\cD_\da$ are realized such that on the tetrad one has
\begin{align}
\acom{\cD_\alpha}{\5\cD_\da}\viel a\mu=-2z\sigma^\nu{}_{\alpha\da}\T{\nu\mu}a,\quad
\acom{\cD_\alpha}{\cD_\beta}\viel a\mu=\acom{\5\cD_\da}{\5\cD_\dbe}\viel a\mu=0
\label{d2}
\end{align}
with $\sigma^\mu{}_{\alpha\da}=\Viel a\mu\sigma^a{}_{\alpha\da}$, and $\T{\mu\nu}a$ as in \eqref{a22},
and on all fields which transform scalarly under general coordinate transformations ($\lambda_\alpha{}^a$, $\5\lambda_\da{}^a$ and auxiliary fields transforming scalarly under general coordinate transformations) one has
\begin{align}
\acom{\cD_\alpha}{\5\cD_\da}=-2z\sigma^a{}_{\alpha\da}\cD_a,\quad
\acom{\cD_\alpha}{\cD_\beta}=\acom{\5\cD_\da}{\5\cD_\dbe}=0
\label{d3}
\end{align}
Hence, we presume that \eqref{d2} and \eqref{d3} hold off-shell, and thus that the commutator algebra of the infinitesimal supersymmetry transformations and general coordinate transformations can be closed off-shell by means of auxiliary fields. It need not be presumed that the auxiliary fields are the components of a vector field even though this of course is the case we have in mind.

\eqref{d2} and \eqref{d3} imply that the commutator of two infinitesimal supersymmetry transformations is an infinitesimal general coordinate transformation, on all fields. Indeed, introducing an anticommuting constant spinor with components $\varepsilon^\alpha$ and the complex conjugated spinor with components $\5\varepsilon^\da$ as parameters of  the infinitesimal global supersymmetry transformations according to
\begin{align}
\delta^{\mathrm{susy}}_\varepsilon =z(\varepsilon^\alpha \cD_\alpha+\5\varepsilon^\da \5\cD_\da),
\label{d4}
\end{align}
one obtains, on all fields, that the commutator of two such supersymmetry transformations with parameters 
$\varepsilon_1^\alpha, \5\varepsilon_1^\da$, and $\varepsilon_2^\alpha, \5\varepsilon_2^\da$, respectively, is an infinitesimal general coordinate transformation $\delta_v$ with parameters $v^\mu=2z^3(\varepsilon_1\sigma^\mu \5\varepsilon_2-
\varepsilon_2\sigma^\mu \5\varepsilon_1)$:
\begin{align}
\com{\delta^{\mathrm{susy}}_{\varepsilon_1}}{\delta^{\mathrm{susy}}_{\varepsilon_2}} =\delta_v,\quad
v^\mu=2z^3(\varepsilon_1\sigma^\mu \5\varepsilon_2-
\varepsilon_2\sigma^\mu \5\varepsilon_1)
\label{d5}
\end{align}

Now, \eqref{d1} through \eqref{d3} imply that on all fields which transform scalarly under general coordinate transformations and are constructible out of the fields $\viel a\mu$, $\lambda_\alpha{}^\mu$, $\5\lambda_\da{}^\mu$, the auxiliary fields and derivatives thereof (such as $\T{ab}c$, $\lambda_\alpha{}^a$, $\5\lambda_\da{}^a$, $B^a$ and covariant derivatives thereof) one has
\begin{align}
[\cD_A,\cD_B\}=-\T {AB}C\cD_C
\label{d6}
\end{align}
where the indices $A$, $B$, $C$ run over Lorentz vector indices $a$ and spinor indices $\alpha, \da$, and
$[\cD_A,\cD_B\}$ denotes the commutator $\com{\cD_A}{\cD_B}$ if $A$ or $B$ is a Lorentz vector index and the anticommutator $\acom{\cD_A}{\cD_B}$ if both $A$ and $B$ are spinor indices:
\begin{align}
[\cD_A,\cD_B\}=\cD_A\cD_B-(-)^{{|A|\, |B|}}\cD_B\cD_A,\quad |a|=0, \ |\alpha|=|\da|=1
\label{d6a}
\end{align}
Using $\com {\6_\mu}{\cD_\alpha}=\com {\6_\mu}{\5\cD_\da}=0$, \eqref{d1} through \eqref{d3} imply
\begin{align}
&\T {\alpha\da}a=\T {\da\alpha}a=2z\sigma^a{}_{\alpha\da},\ 
\T {\alpha a}b=-\T {a\alpha}b=\Viel a\mu\cD_\alpha \viel b\mu,
\nonumber\\
&\T {\da a}b=-\T {a\da}b=\Viel a\mu\5\cD_\da \viel b\mu,\
\T {ab}c=\Viel a\mu\Viel b\nu\T {\mu\nu}c
\label{d7}
\end{align}
The Bianchi identities following from \eqref{d6} are
\begin{align}
\csum {ABC}{36}(-)^{|A|\, |C|}
     (\cD_A\T {BC}D+\T {AB}E\T {EC}D)=0
\label{d8}
\end{align}
where ${\displaystyle \csum {}{25}}X_{ABC}=X_{ABC}+X_{BCA}+X_{CAB}$ denotes the cyclic sum. The Bianchi identities \eqref{d8} for the torsions in \eqref{d7} actually are implied by \eqref{d2}, i.e. these Bianchi identities hold whenever equations \eqref{d2} hold, as can be checked readily.

Now we assign ``mass dimensions'', denoted by $[\ ]$, to the various fields and operators according to
\begin{align}
[\viel a\mu]=0,\ [\lambda_\alpha{}^a]=[\cD_\alpha]=1/2,\ [\6_\mu]=1
\label{d9}
\end{align}
Presuming that the supersymmetry transformations of $\viel a\mu$ only contain parameters with vanishing mass dimension and taking into account the Lorentz indices, $\cD_\alpha \viel a\mu$ must be linear in $\5\lambda_\da{}^a$. Hence, in order to construct supersymmetric extensions of actions with Lagrangians \eqref{a24} corresponding to supersymmetric free field theories with coefficients $b_i$ as in \eqref{b34}, we now shall consider
\begin{align}
\cD_\alpha\viel a\mu=[b_1 \sigma_\mu\5\lambda^a+(1-b_1) \sigma^a\5\lambda_\mu+
(b_1-1)\viel a\mu\sigma^b\5\lambda_b+\Ii (1-b_1)\epsilon_\mu{}^{abc}\sigma_b\5\lambda_c]_\alpha
\label{d10}
\end{align}
In addition we shall need $\cD_\alpha \5\lambda_\da{}^a$. In order to derive it, we 
decompose $\5\lambda_\da{}^a$ into a spin-1/2 part $\chi$ and a spin-3/2 part $\5\psi$ according to
\begin{align}
\5\lambda_{\da\beta\dbe}\equiv
\5\lambda_\da{}^a\sigma_{a\beta\dbe}=\tfrac 12\epsilon_{\da\dbe}\chi_\beta+\5\psi_{\da\dbe\beta},\quad
\5\psi_{\da\dbe\beta}=\5\psi_{\dbe\da\beta}
\label{d11}
\end{align}
which implies
\begin{align}
\chi_\alpha=(\sigma^a\5\lambda_a)_\alpha,\
\5\psi_{\da\dbe\alpha}=\sigma^a{}_{\alpha(\da}\5\lambda_{\dbe)a},\
\5\lambda_{\da a}=-\tfrac 14(\chi\sigma_a)_\da+\tfrac 12\5\sigma_a{}^{\dbe\beta}\5\psi_{\da\dbe\beta}
\label{d11a}
\end{align}
\eqref{b3}, \eqref{d9} and the Bianchi identity \eqref{d8} with indices $ABCD=\alpha\beta c d$ suggest that $\cD_\alpha \5\lambda_\da{}^a$ is bilinear in $\5\lambda$. Therefore we make the following general Ans\"atze for 
$\cD_\alpha \chi_\beta$ and $\cD_\alpha\5\psi^{\da\dbe}{}_\beta$:
\begin{align}
\cD_\alpha \chi_\beta=\,&\epsilon_{\alpha\beta}(d_1\chi\chi+d_2\5\psi\5\psi)
\label{d12}\\
\cD_\alpha\5\psi^{\da\dbe}{}_\beta=\,&d_3\epsilon_{\alpha\beta}\5\psi^{\da\dbe\gamma}\chi_\gamma
+d_4\5\psi^{\da\dbe}{}_{(\alpha}\chi_{\beta)}+d_5\5\psi_\dga{}^{(\da}{}_\alpha\5\psi^{\dbe)\dga}{}_\beta
\label{d13}
\end{align}
with $\chi\chi=\chi^\alpha\chi_\alpha$, $\5\psi\5\psi=\5\psi_{\da\dbe}{}^\alpha\5\psi^{\da\dbe}{}_\alpha$, and (in general complex) coefficients $d_1,\dots,d_5$. Using \eqref{d7} with $\cD_\alpha\viel a\mu$ as in \eqref{d10}, and \eqref{d12} and \eqref{d13} in the Bianchi identity \eqref{d8} with indices $ABCD=\alpha\beta c d$ yields
\begin{align}
d_1=\frac{3-4b_1}{4},\ d_2=\frac{1}{2(4b_1-3)},\ d_3=\frac{3-4b_1}{2},\ d_4=0,\ d_5=-1
\label{d14}
\end{align}
We note that \eqref{d12} through \eqref{d14} also imply $\acom{\cD_\alpha}{\cD_\beta}\5\lambda_\da{}^a=0$,
as well as
\begin{align}
\cD_\alpha(\chi\chi)=\frac{1}{3-4b_1}\, \chi_\alpha \5\psi\5\psi,\qd 
\cD_\alpha(\5\psi\5\psi)=\frac {3-4b_1}2\,\chi_\alpha \5\psi\5\psi
\label{d15}
\end{align}
Furthermore we note that 
\begin{align}
\chi\chi=-\5\lambda_a\5\lambda^a-2\5\lambda^a\5\sigma_{ab}\5\lambda^b,\qd 
\5\psi\5\psi=-\tfrac 32\5\lambda_a\5\lambda^a+\5\lambda^a\5\sigma_{ab}\5\lambda^b
\label{d15a}
\end{align}

In appendix \ref{brst} it is shown that, under the above prerequisites, Lagrangians which are invariant, up to total divergences, under general coordinate transformations and supersymmetry transformations generated by \eqref{d4} can be constructed by means of 
operators $\cP$ and $\5\cP$ defined according to
\begin{align}
\cP =\,&\cD^2+2(4b_1-3)\chi\cD+\tfrac 12(4b_1-3)^2\chi\chi+\5\psi\5\psi
\label{d16}\\
\5\cP= \,&-z^2 [\5\cD^2+2(4b_1-3)\5\chi\5\cD+\tfrac 12(4b_1-3)^2\5\chi\5\chi+\psi\psi]
\label{d17}
\end{align}
with $\cD^2=\cD^\alpha\cD_\alpha$, $\chi\cD=\chi^\alpha\cD_\alpha$, $\5\cD^2=\5\cD_\da\5\cD^\da$,
$\5\chi\5\cD=\5\chi_\da\5\cD^\da$, $\5\chi\5\chi=\5\chi_\da\5\chi^\da$, 
$\psi\psi=\psi^{\alpha\beta}{}_\da\psi_{\alpha\beta}{}^\da$, and $\chi\chi$ and $\5\psi\5\psi$ as above.
We note that
\begin{align}
\tfrac 12(4b_1-3)^2\chi\chi+\5\psi\5\psi=
-\tfrac 12[(4b_1-3)^2+3]\5\lambda_a\5\lambda^a
-16(b_1-1)(b_1-\tfrac 12)\5\lambda^a\5\sigma_{ab}\5\lambda^b
\label{d16a}
\end{align}

The operator $\cP$ and the field polynomial $\tfrac 12(4b_1-3)^2\chi\chi+\5\psi\5\psi$ fulfill the following equations:
\begin{align}
[\cD_\alpha+(4b_1-3)\chi_\alpha]\cP =0,\
[\cD_\alpha+(4b_1-3)\chi_\alpha](\tfrac 12(4b_1-3)^2\chi\chi+\5\psi\5\psi)=0
\label{d18}
\end{align}
where the first equation in \eqref{d18} is an operator identity.
Accordingly $\5\cP$ and $\tfrac 12(4b_1-3)^2\5\chi\5\chi+\psi\psi$ fulfill
\begin{align}
[\5\cD_\da+(4b_1-3)\5\chi_\da]\5\cP =0,\
[\5\cD_\da+(4b_1-3)\5\chi_\da](\tfrac 12(4b_1-3)^2\5\chi\5\chi+\psi\psi)=0
\label{d19}
\end{align}
In appendix \ref{brst} it is shown that, under the above prerequisites, the ``density formula'' 
\begin{align}
e\5\cP h(\cT)+c.c.
\label{d20}
\end{align}
provides a function of the fields and their derivatives which is invariant, up to total divergences, under general coordinate transformations and supersymmetry transformations generated by \eqref{d4} for \textit{any} function $h(\cT)$ of the tensors $\cT^r$ given in \eqref{brs12} (with $\phi^M$ as in \eqref{brs3} where $B^a$ may be replaced by other auxiliary fields) which  fulfills
\begin{align}
[\cD_\alpha+(4b_1-3)\chi_\alpha]h(\cT)=0
\label{d21}
\end{align}
I remark that $(4b_1-3)\chi_\alpha h(\cT)$ in \eqref{d21} originates from the presence of the torsion $\T{\alpha a}b$ in \eqref{d6}, as the analysis in appendix \ref{brst} shows.

Using the mass dimensions given in \eqref{d9} one infers that the supersymmetric extensions of actions with Lagrangians \eqref{a24} arise from functions $h(\cT)$ which are quadratic in the $\lambda^{\alpha a}$ and $\5\lambda^{\da a}$. Furthermore these functions  must be Lorentz invariant. According to \eqref{b2} (with $b_9=-1$) one always has 
$\cD_\alpha\lambda^{\beta a}=-z\sigma^{bc}{}_\alpha{}^\beta\T {bc}a+\dots$ This implies that the 
supersymmetric extensions of actions with Lagrangians \eqref{a24} arise from Lorentz invariant functions
$h(\cT)$ which are quadratic in the  $\5\lambda^{\da a}$, and thus that any such function is a linear combination (with complex coefficients) of $\chi\chi$ and $\5\psi\5\psi$. \eqref{d15} implies that the only linear combinations of $\chi\chi$ and $\5\psi\5\psi$ which fulfill \eqref{d21} are
$a(\tfrac 12(4b_1-3)^2\chi\chi+\5\psi\5\psi)$ where $a$ is an arbitrary complex number. This implies that (in the generic cases)
the supersymmetric extensions of actions with Lagrangians \eqref{a24} arise from
\begin{align}
a e \5\cP\,[\tfrac 12(4b_1-3)^2\chi\chi+\5\psi\5\psi]+c.c.,\qd a\in\mathbb{C}
\label{d22}
\end{align}
Now, in general \eqref{d22} provides two contributions to the Lagrangian which are separately invariant, up to total divergences, under general coordinate transformations and supersymmetry transformations and are proportional to the real and the imaginary part of $e \5\cP\,[\tfrac 12(4b_1-3)^2\chi\chi+\5\psi\5\psi]$, respectively. 
In other words, any Lagrangian \eqref{d22} involves at most two arbitrary real coefficients.
Furthermore it can and does happen (cf. section \ref{ex}) that the Lagrangian \eqref{d22}, up to a total divergence, actually involves only one arbitrary real coefficient, namely when the real or the imaginary part of $e \5\cP\,[\tfrac 12(4b_1-3)^2\chi\chi+\5\psi\5\psi]$ is a total divergence. Then that coefficient is just an overall factor of the Lagrangian, i.e. the Lagrangian essentially is unique. Notice that this also implies that at most two of the coefficients $a_1,\dots,a_4$ in the Lagrangian can be independent.
This shows that supersymmetric ``teleparallel'' theories are much more constrained than non-supersymmetric ones.

I remark that the analysis in appendix \ref{brst} analogously can be conducted for the supersymmetric free theories of section \ref{susyfree} (presuming that the commutator algebra of the supersymmetry transformations and spacetime translations can be closed off-shell, up to gauge transformations, by means of auxiliary fields). In place of \eqref{d20} and \eqref{d21} one then obtains $-z^2\5D^2\4h(\4T)$ where
$\4h(\4T)$ is any function of $\F {\mu\nu}\rho$, $\lambda^{\alpha\mu}$, $\5\lambda^{\da\mu}$, $B^\mu$ (or other auxiliary fields) and derivatives thereof which fulfills $D_\alpha \4h(\4T)=0$. According to \eqref{b3} one has $D_\alpha\5\lambda^{\da\mu}=0$ which implies that $\chi\chi$ and $\5\psi\5\psi$ fulfill
$D_\alpha(\chi\chi)=0$ and $D_\alpha(\5\psi\5\psi)=0$. The analog of \eqref{d22} in the supersymmetric free field theories thus reads $-z^2\5D^2(a\chi\chi+b\5\psi\5\psi)+c.c.$ where $a$ and $b$ are arbitrary complex coefficients.\footnote{This implies that a Lagrangian of a supersymmetric free theory corresponding to \eqref{a2} may have up to four arbitrary real coefficients, just as \eqref{a2}. However it turns out that there is a 
linear combination of $\5D^2(\chi\chi)$ and $\5D^2(\5\psi\5\psi)$
which is a total divergence which then reduces the number of arbitrary real coefficients to at most three.} This shows that and explains why supersymmetric ``teleparallel'' theories are more constrained than the corresponding supersymmetric free field theories: the reason is the condition \eqref{d21} in combination with the nonlinear terms in the supersymmetry transformations of the fields.

Lagrangians of supersymmetric actions more general than \eqref{d22} arise from
\begin{align}
e \5\cP\cP f(\cT)+c.c.
\label{d23}
\end{align}
where $f(\cT)$ is any function of the tensors $\cT^r$ (notice that $h(\cT)=\cP f(\cT)$ fulfills \eqref{d21} for any $f(\cT)$ due to the first equation in \eqref{d18}). A constant contribution to $f$ provides again \eqref{d22}. Non-constant contributions to $f$ provide supersymmetric actions containing higher powers of the torsion $\T{ab}c$ and/or terms with more than two derivatives.

\textit{Remark:} One has
\begin{align}
\cD_\alpha e =(4b_1-3)e\chi_\alpha, \qd
\cD^2 e =\tfrac 12(4b_1-3)^2e\chi\chi+e\5\psi\5\psi
\label{d24}
\end{align}
This implies, for any function $f(\cT)$,
\begin{align}
e\cP f(\cT)=\cD^2[e f(\cT)],\ 
e\5\cP f(\cT)=-z^2\5\cD^2[e f(\cT)],\ 
e \5\cP\cP f(\cT)=-z^2\5\cD^2\cD^2[ef(\cT)]
\label{d25}
\end{align}
Furthermore, if $h(\cT)$ fulfills \eqref{d21} then one has $\cD_\alpha[eh(\cT)]=0$. 
This makes it obvious that \eqref{d20} with $h(\cT)$ fulfilling \eqref{d21}, and \eqref{d23} provide contributions to supersymmetric actions.

\subsection{Lagrangians with further supersymmetry multiplets}\label{matter}

We now discuss the construction of supersymmetric actions involving further supersymmetry multiplets. We first introduce super Yang-Mills multiplets whose fields are Yang-Mills gauge fields $A_\mu{}^i$, gauginos $\lambda_\alpha{}^i$ (and their complex conjugates $\5\lambda_\da{}^i$)  and real scalar auxiliary fields $D^i$ where $i$ enumerates generators $\delta_i$ of a reductive Lie algebra. We proceed analogously to section \ref{tele}: we extend the covariant derivatives \eqref{d1} according to
\begin{align}
\cD_a=\Viel a\mu(\6_\mu-A_\mu{}^i\delta_i),
\label{d30}
\end{align}
impose
\begin{align}
\cD_\alpha A_\mu{}^i=(\sigma_\mu\5\lambda^i)_\alpha,
\label{d31}
\end{align}
and shall presume that the supersymmetry transformations $\cD_\alpha$ and $\5\cD_\da$
on $A_\mu{}^i$ fulfill the algebra
\begin{align}
\acom{\cD_\alpha}{\5\cD_\da}A_\mu{}^i=-2z\sigma^\nu{}_{\alpha\da}\F{\nu\mu}i,\quad
\acom{\cD_\alpha}{\cD_\beta}A_\mu{}^i=\acom{\5\cD_\da}{\5\cD_\dbe}A_\mu{}^i=0
\label{d32}
\end{align}
where $\F{\mu\nu}i$ are the components of the Yang-Mills field strengths:
\begin{align}
\F{\mu\nu}i=\6_\mu A_\nu{}^i-\6_\nu A_\mu{}^i+\f jki A_\mu{}^jA_\nu{}^k
\label{d33}
\end{align}
where $\f ijk$ are the structure constants of the Lie algebra of the $\delta_i$,
\begin{align}
\com{\delta_i}{\delta_j}=\f ijk\delta_k
\label{d33a}
\end{align}
Furthermore we presume that the supersymmetry transformations $\cD_\alpha$ and $\5\cD_\da$ are realized off-shell such that, in place of \eqref{d6}, one has, on the fields \eqref{brs36} (where $B^a$ may be replaced with other auxiliary fields transforming scalarly under general coordinate transformations) and their covariant derivatives,
\begin{align}
[\cD_A,\cD_B\}=-\F {AB}i\delta_i-\T {AB}C\cD_C
\label{d36}
\end{align}
with nonvanishing field strengths
\begin{align}
\F {\alpha a}i=-\F {a\alpha}i=(\sigma_a\5\lambda^i)_\alpha,\ \F {\da a}i=-\F {a \da}i=(\lambda^i\sigma_a)_\da,\
\F {ab}i=\Viel a\mu\Viel b\nu\F{\mu\nu}i
\label{d37}
\end{align}
\eqref{d36} gives the additional Bianchi identities
\begin{align}
\csum {ABC}{36}(-)^{|A|\, |C|}
     (\cD_A\F {BC}i+\T {AB}D\F {DC}i)=0
\label{d38}
\end{align}
In place of \eqref{d5} one now gets
\begin{align}
\com{\delta^{\mathrm{susy}}_{\varepsilon_1}}{\delta^{\mathrm{susy}}_{\varepsilon_2}} =\delta_v+\delta^{\mathrm{YM}}_w,\quad
w^i=-2z^3(\varepsilon_1\sigma^\mu \5\varepsilon_2-
\varepsilon_2\sigma^\mu \5\varepsilon_1)A_\mu{}^i
\label{d35}
\end{align}
where $\delta_v$ denotes an infinitesimal general coordinate transformation with parameters $v^\mu$ as in \eqref{d5}, and
$\delta^{\mathrm{YM}}_w$ denotes an infinitesimal Yang-Mills gauge transformation with gauge parameters $w^i$ which reads on $A_\mu{}^i$, $\lambda_\alpha{}^i$ and $D^i$, respectively:
\begin{align}
\delta^{\mathrm{YM}}_w A_\mu{}^i=\6_\mu w^i+\f jki w^k A_\mu{}^j,\
\delta^{\mathrm{YM}}_w \lambda_\alpha{}^i=\f jki w^k\lambda_\alpha{}^j,\
\delta^{\mathrm{YM}}_w D^i=\f jki w^k D^j
\label{d35a}
\end{align}

Now, in order to construct, along the lines of the section \ref{tele}, supersymmetric actions containing the fields of the super Yang-Mills multiplets we need $\cD_\alpha\5\lambda_\da{}^i$. We make the following Ansatz:
\begin{align}
\cD_\alpha \5\lambda_\da{}^i=\,&d_6\5\lambda_\da{}^a(\sigma_a\5\lambda^i)_\alpha
+d_7(\sigma_a\5\lambda^a)_\alpha\5\lambda_\da{}^i
\label{d42}
\end{align}
Using \eqref{d37}, \eqref{d42}, and \eqref{d7} with $\cD_\alpha\viel a\mu$ as in \eqref{d10} in the Bianchi identity \eqref{d38} with indices $ABC=\alpha\beta c$ yields
\begin{align}
d_6=-1,\ d_7=2(1-b_1)
\label{d44}
\end{align}
We note that \eqref{d31}, \eqref{d42} and \eqref{d44} imply $\acom{\cD_\alpha}{\cD_\beta}A_\mu{}^i=0$, i.e. \eqref{d42} with $d_6$ and $d_7$ as in \eqref{d44} is consistent with \eqref{d32}.

Now it is straightforward to verify that
\begin{align}
[\cD_\alpha+(4b_1-3)\chi_\alpha](d_{ij}\5\lambda^i\5\lambda^j)=0
\label{d48}
\end{align}
where $d_{ij}$ are constant components of a symmetric invariant tensor of the Lie algebra of the $\delta_i$, i.e. 
\begin{align}
\forall i,j,k:\ \f kim d_{m j}+\f kjm d_{im}=0
\label{d48a}
\end{align}
According to \eqref{d48} and \eqref{d48a}, $h=d_{ij}\5\lambda^i\5\lambda^j$ fulfills both \eqref{d21} and \eqref{brs42}, and thus \eqref{d20} provides a contribution to the Lagrangian of a supersymmetric action given by
\begin{align}
be \5\cP\,(d_{ij}\5\lambda^i\5\lambda^j)+c.c.
\label{d52}
\end{align}
where $b$ is an arbitrary complex number. Hence, in general \eqref{d52} provides two contributions to the Lagrangian which are separately invariant, up to total divergences, under general coordinate transformations and supersymmetry transformations and are proportional to the real and the imaginary part of $e \5\cP\,(d_{ij}\5\lambda^i\5\lambda^j)$, respectively.

Using the Bianchi identity \eqref{d38} with indices $ABC=\alpha\dbe c$, one infers
that \eqref{d52} provides terms quadratic in the Yang-Mills field strengths
 $\F{\mu\nu}i$, for one obtains
 \begin{align}
\cD_\alpha\lambda^{\beta i}=-z\sigma^{ab}{}_\alpha{}^\beta\F{ab}i+\ldots,\ 
\5\cD_\da\5\lambda^{\dbe i}=z\5\sigma^{ab\dbe}{}_\da\F{ab}i+\ldots,
\label{d52a}
\end{align}
Contributions to a supersymmetric action with a Lagrangian containing higher powers of the Yang-Mills field strengths, terms with more than two spacetime derivatives, and/or terms containing both the torsion $\T{\mu\nu}a$ and the Yang-Mills field strengths arise from
\eqref{d23}.

Next we shall discuss the inclusion of scalar supersymmetry multiplets whose lowest component fields are complex scalar fields $\varphi^m$ which transform under Yang-Mills gauge transformations according to
 \begin{align}
\delta^{\mathrm{YM}}_w\varphi^m=-w^i T_i{}^m{}_n\varphi^n
\label{d53}
\end{align}
where $T_i{}^m{}_n$ are the entries of matrices $T_i$ representing the $\delta_i$ according to
$\com{T_i}{T_j}=\f ijk T_k$.
We denote the higher component fields of scalar supersymmetry multiplets by $\eta_\alpha{}^m$ and $F^m$ where
$\eta^m$ are complex spinor fields and $F^m$ are complex scalar auxiliary fields.

For the supersymmetry transformations $\cD_\alpha$ of $\varphi^m$, $\eta_\alpha{}^m$ and $F^m$ we use, as usual,
\begin{align}
\cD_\alpha\varphi^m=\eta_\alpha{}^m,\ 
\cD_\alpha\eta_\beta{}^m=\epsilon_{\beta\alpha}F^m,\ 
\cD_\alpha F^m=0
\label{d55}
\end{align}

Let us now discuss the transformations $\cD_\alpha$ of the complex conjugated fields whose components are denoted by
$\5\varphi^{\5m}$, $\5\eta_\da{}^{\5m}$ and $\5F^{\5m}$.
Usually (in standard supergravity) one imposes the antichirality condition $\cD_\alpha\5\varphi^{\5m}=0$. This can be done here too. However, imposing $\cD_\alpha\5\varphi^{\5m}=0$ for all scalar multiplets does not allow the construction of a superpotential for the scalar multiplets. Namely a superpotential would arise from \eqref{d20} for a function $h(\5\varphi)$ of the undifferentiated fields $\5\varphi$. Now, this function must fulfill \eqref{d21} which is not the usual antichirality condition $\cD_\alpha h(\5\varphi)=0$ but the
condition $\cD_\alpha h(\5\varphi)=(3-4b_1)\chi_\alpha h(\5\varphi)$ which cannot be fulfilled nontrivially when all $k^{\5m}$ are zero (recall that $b_1\neq 3/4$). This suggests to relax the antichirality condition
$\cD_\alpha\5\varphi^{\5m}=0$ and use instead of it the following $\cD_\alpha$-transformations:
\begin{align}
\cD_\alpha\5\varphi^{\5m}=\chi_\alpha k^{\5m} \5\varphi^{\5m}\quad (\mathrm{no\ sum\ over}\ \5m)
\label{d56}
\end{align}
where $k^{\5m}$ are numbers which coincide for the component fields of any particular scalar supersymmetry multiplet (whose component fields transform under Yang-Mills gauge transformations according to some irreducible representation $\{T_i\}$ of the Lie algebra of the $\delta_i$) but may differ for different such multiplets. One now easily derives $\cD_\alpha\5\eta_\da{}^{\5m}$ and 
$\cD_\alpha\5F^{\5m}$ by imposing and using the algebra \eqref{d36}. The result is not spelled out here.
Then \eqref{d20} through \eqref{d23} can be used to construct contributions to supersymmetric actions containing the fields of the scalar multiplets and their derivatives.

\subsection{Lagrangians for $b_1=1$}\label{ex}

In this section we present the nonlinear extensions of the supersymmetric free field theories provided in section \ref{susyfree} for $b_1=1$. The nonlinear extensions of the supersymmetry transformations given in \eqref{b1aa} through \eqref{b4aa} turn out to be
\begin{align}
b_1=1:\qd &\cD_\alpha\viel a\mu=(\sigma_\mu\5\lambda^a)_\alpha
\label{b1bb}\\
&\cD_\alpha\lambda^{\beta\mu}=z(\Ii\delta_\alpha^\beta  \7B^{\mu}
-\sigma^{\nu\rho}{}_\alpha{}^\beta \T{\nu\rho}\mu)
\label{b2bb}\\
&\cD_\alpha\5\lambda^{\da\mu}=0
\label{b3bb}\\
&\cD_\alpha \7B^{\mu}=-\Ii\sigma^\nu{}_{\alpha\da} \6_\nu\5\lambda^{\da\mu}
+\Ii\5\lambda^{\da\nu}\6_\nu\sigma^\mu{}_{\alpha\da}
\label{b4bb}
\end{align}
It is important that the index $\mu$ of $\lambda^{\beta\mu}$, $\5\lambda^{\da\mu}$ and $\7B^{\mu}$ in \eqref{b2bb} through \eqref{b4bb} is a contravariant (i.e. upper) world index. The corresponding transformations of these fields with a covariant (i.e. lower) index $\mu$ are more complicated, as are the transformations with a Lorentz vector index in place of a contravariant world index. These transformations are obtained from \eqref{b1bb} through \eqref{b4bb} as usual, using
$\cD_\alpha \lambda^{\beta}{}_\mu=\cD_\alpha(g_{\mu\nu}\lambda^{\beta\nu})$ with 
$g_{\mu\nu}=\viel a\mu\viel b\nu\eta_{ab}$, 
$\cD_\alpha \lambda^{\beta a}=\cD_\alpha(\viel a\mu\lambda^{\beta \mu})$ etc. The resultant transformations of $\lambda^{\beta a}$, $\5\lambda^{\da a}$ and $\7B^{a}$ are
\begin{align}
&\cD_\alpha\lambda^{\beta a}=z(\Ii\delta_\alpha^\beta  \7B^{a}
-\sigma^{bc}{}_\alpha{}^\beta \T{bc}a)-\lambda^{\beta b}(\sigma_b\5\lambda^a)_\alpha
\label{b5bb}\\
&\cD_\alpha\5\lambda^{\da a}=-\5\lambda^{\da b}(\sigma_b\5\lambda^a)_\alpha
\label{b6bb}\\
&\cD_\alpha \7B^a=
-\Ii(\sigma^b\cD_b\5\lambda^a)_\alpha
+\Ii \T {bc}a(\sigma^b\5\lambda^c)_\alpha
+\7B^b(\sigma_b\5\lambda^a)_\alpha 
\label{b7bb}
\end{align}
\eqref{b1bb} through \eqref{b4bb} (and the complex conjugates of these transformations) imply that on all fields $\viel a\mu$, $\lambda^{\beta\mu}$, $\5\lambda^{\da\mu}$, $\7B^\mu$ the commutator of two infinitesimal supersymmetry transformations \eqref{d4} fulfills \eqref{d5} off-shell, with $\delta_v$ a standard infinitesimal general coordinate transformation which acts, e.g., on $\viel a\mu$ and $\lambda^{\beta\mu}$ according to
\begin{align}
\delta_v\viel a\mu=v^\nu\6_\nu\viel a\mu+(\6_\mu v^\nu)\viel a\nu,\qd
\delta_v\lambda^{\beta\mu}=v^\nu\6_\nu\lambda^{\beta\mu}-(\6_\nu v^\mu)\lambda^{\beta\nu}
\label{b8bb}
\end{align}
Accordingly the $\cD_\alpha$ and $\5\cD_\da$ fulfill the algebra \eqref{d3} off-shell on 
$\T {ab}c$, $\lambda^{\beta a}$, $\5\lambda^{\da a}$, $\7B^{a}$ and covariant derivatives thereof which was used in the derivation of the Lagrangians presented in \eqref{d20} through \eqref{d23}, cf. appendix \ref{brst}.
\eqref{d22} thus gives a Lagrangian of a supersymmetric action in the case $b_1=1$. In this case 
\eqref{d16a} and its complex conjugate and \eqref{d17} give:
\begin{align}
b_1=1:\qd &\tfrac 12(4b_1-3)^2\chi\chi+\5\psi\5\psi=
-2\5\lambda_a\5\lambda^a
\label{b9bb1}
\\
&\tfrac 12(4b_1-3)^2\5\chi\5\chi+\psi\psi=
-2\lambda_a\lambda^a
\label{b9bb2}
\\
 &\5\cP=-z^2 (\5\cD^2+2\lambda^a\sigma_a\5\cD-2\lambda_a\lambda^a)
\label{b9bb}
\end{align}
Using \eqref{b5bb} through \eqref{b7bb} (and the complex conjugates of these transformations) one obtains that the imaginary part of $e\5\cP(\5\lambda_a\5\lambda^a)$ is a total divergence (one gets 
$\tfrac 12 e\5\cP(\5\lambda_a\5\lambda^a)-c.c.=
-\Ii e\epsilon^{abcd}T_{abe}\T {cd}e+\ldots=\6_\mu (-2\Ii e\epsilon^{\mu\nu\rho\sigma}e_{\nu a}\T {\rho\sigma}a+\dots$)).
The real part of $e\5\cP(\5\lambda_a\5\lambda^a)$ is
\begin{align}
& -\tfrac 12 z^2 e(\5\cD^2+2\lambda^a\sigma_a\5\cD-2\lambda_a\lambda^a)
(\5\lambda_b\5\lambda^b)+c.c.
\nonumber\\
&\,=e(2T_{abc}T^{abc}
-4z^3\5\lambda^a\5\sigma^b\cD_b\lambda_a-4z^3\lambda^a\sigma^b\cD_b\5\lambda_a
-4\7B_a\7B^a+\ldots)
\label{b10bb}
\end{align}
where the ellipses denote terms which are at least trilinear in the
$\T {ab}c$, $\lambda^{\beta a}$, $\5\lambda^{\da a}$ and $\7B^{a}$.  I stress that the analysis of appendix \ref{brst} implies that \eqref{b10bb}, up to a total divergence and an overall factor, is the unique counterpart of a Lagrangian \eqref{a24} in a supersymmetric theory in the case $b_1=1$, i.e. a theory with Lagrangian \eqref{a24} has a supersymmetric  extension with $b_1=1$ only for $a^\prime_2=a_3=a^\prime_4=0$.

Finally we provide for the case $b_1=1$ the supersymmetry transformations \eqref{d4} of the component fields of super Yang-Mills multiplets which fulfill \eqref{d35} off-shell. These are generated by the following transformations 
$\cD_\alpha$ and the corresponding complex conjugated transformations $\5\cD_\da$:
\begin{align}
b_1=1:\qd&\cD_\alpha A_\mu{}^i=(\sigma_\mu\5\lambda^i)_\alpha
\label{b11bb}\\
&\cD_\alpha\lambda^{\beta i}=z(\Ii\delta_\alpha^\beta  D^i
-\sigma^{bc}{}_\alpha{}^\beta \F{bc}i)-\lambda^{\beta b}(\sigma_b\5\lambda^i)_\alpha
\label{b12bb}\\
&\cD_\alpha\5\lambda^{\da i}=-\5\lambda^{\da b}(\sigma_b\5\lambda^i)_\alpha
\label{b13bb}\\
&\cD_\alpha D^i=
-\Ii(\sigma^b\cD_b\5\lambda^i)_\alpha
+\Ii \F {bc}i(\sigma^b\5\lambda^c)_\alpha
+\7B^b(\sigma_b\5\lambda^i)_\alpha 
\label{b14bb}
\end{align}

\eqref{d52} provides a contribution to the Lagrangian of supersymmetric actions involving the component fields of super Yang-Mills multiplets. As the imaginary part of $e\5\cP(\5\lambda_a\5\lambda^a)$, the imaginary part of $e\5\cP(d_{ij}\5\lambda^i\5\lambda^j)$ is a total divergence. The real part of $e\5\cP(d_{ij}\5\lambda^i\5\lambda^j)$ is
\begin{align}
&-\tfrac 12 z^2e(\5\cD^2+2\lambda^a\sigma_a\5\cD-2\lambda_a\lambda^a)
(d_{ij}\5\lambda^i\5\lambda^j)+c.c.
\nonumber\\
&\,=ed_{ij}(2\F {ab}i F^{abj}
-4z^3\5\lambda^i\5\sigma^b\cD_b\lambda^j-4z^3\lambda^i\sigma^b\cD_b\5\lambda^j
-4 D^i D^j+\ldots)
\label{b15bb}
\end{align}

Concerning the construction of supersymmetric actions with Lagrangians containing terms with the component fields of scalar multiplets, higher powers of the torsion and/or Yang-Mills field strengths, and/or terms with more than two derivatives we refer to sections \ref{tele} and \ref{matter}.

\mysection{Addendum}\label{add}

In this addendum it is shown that every supersymmetric free field theory with a Lagrangian \eqref{b5}, with $\LB$ as in \eqref{a2} excluding the cases \eqref{a9}, and supersymmetry transformations \eqref{b1} through \eqref{b3} fulfilling \eqref{b4} is a free field theory with parameters \eqref{sol1} or can be obtained from one of these free field theories by a local field redefinition.\footnote{In particular this applies to the free field theories with parameters \eqref{sol2} or \eqref{sol3}, i.e. each of these free field theories can be obtained by a local field redefinition \eqref{add1} from a free field theory with parameters \eqref{sol1}, as may be verified explicitly.}

Concerning the cases \eqref{a9} we recall once again that the Lagrangian \eqref{a24} is for 
$a_2=-2a_1$, $a_3=-4a_1$, $a_4=0$
($\Leftrightarrow$ $a^\prime_2=4a^\prime_1$,
$a_3=-2a^\prime_1$, $a^\prime_4=0$) proportional to the Einstein-Hilbert Lagrangian up to a total divergence and thus its supersymmetrization is (equivalent to) standard supergravity,  cf. also footnote \ref{sugra}. For other parameters fulfilling \eqref{a9} the Lagrangian \eqref{b5} cannot be
supersymmetric as a mere counting of degrees of freedom shows: in these cases $A_{\mu\nu}$ always provides an odd number of bosonic degrees on-shell according to the usual counting (one degree contributed by $B_{\mu\nu}$, two degrees contributed by $H_{\mu\nu}$, if present) and thus the numbers of bosonic and fermionic degrees of freedom differ on-shell.

Accordingly the theories presented in section \ref{ex} exhaust, modulo local field redefinitions, all supersymmetric theories  
in the
scope of investigation of this paper different from standard supergravity, i.e. any other theory (with $b_1\neq 1$) in this scope different from standard supergravity can be obtained from a theory presented in section \ref{ex} by a local field redefinition.

The result on the free field theories announced above is obtained by considering local redefinitions of the fermionic fields 
which read, suppressing spinor indices,
\begin{align}
\lambda^{\prime}_\mu=b^\prime\lambda_\mu
+a^\prime\lambda^{\nu}\sigma_{\nu\mu},\qd
\lambda_\mu=b\lambda^\prime_\mu
+a\lambda^{\prime\, \nu}\sigma_{\nu\mu}
\label{add1}
\end{align}
where $a$, $b$, $a^\prime$, $b^\prime$ are complex parameters which fufill
\begin{align}
\tfrac 34 a a^\prime+bb^\prime=1,\qd a a^\prime+ab^\prime+ba^\prime=0
\label{add2}
\end{align}
Applying 
the field redefinitions \eqref{add1} to the supersymmetry transformations \eqref{b1} through \eqref{b3} gives
\begin{align}
\5D_\da A_{\mu\nu}=\,&
(\5b_1\lambda_\nu\sigma_\mu+\5b_2\lambda_\mu\sigma_\nu
+\5b_3\eta_{\mu\nu}\lambda_\rho\sigma^\rho
+\5b_4\epsilon_{\mu\nu\rho\sigma}\lambda^\sigma\sigma^\rho)_\da
\nonumber\\
=\,&
[(-\tfrac 12 a \5b_2-\Ii a \5b_4+b\5b_1)\lambda^\prime{}_\nu\sigma_\mu
+(-\tfrac 12 a \5b_1+\Ii a \5b_4+b\5b_2)\lambda^\prime{}_\mu\sigma_\nu
\nonumber\\
&+(\tfrac 12 a\5b_1+\tfrac 12 a\5b_2+\tfrac 32 a \5b_3+b\5b_3)\eta_{\mu\nu}\lambda^\prime{}_\rho\sigma^\rho
\nonumber\\
&+(\tfrac 12 \Ii a\5b_1-\tfrac 12 \Ii a \5b_2+\tfrac 12 a \5b_4+b\5b_4)\epsilon_{\mu\nu\rho\sigma}\lambda^{\prime\,\sigma}\sigma^\rho)]_\da
\nonumber\\
\equiv\,&(\5b^\prime_1\lambda^\prime{}_\nu\sigma_\mu+\5b^\prime_2\lambda^\prime{}_\mu\sigma_\nu
+\5b^\prime_3\eta_{\mu\nu}\lambda^\prime{}_\rho\sigma^\rho
+\5b^\prime_4\epsilon_{\mu\nu\rho\sigma}\lambda^{\prime\,\sigma}\sigma^\rho)_\da
\label{add3}\\
z^{-1}D_\alpha \lambda^{\prime\,\beta}{}_\mu=\,&
z^{-1}D_\alpha (b^\prime \lambda_\mu
+a^\prime \lambda^{\nu}\sigma_{\nu\mu})^\beta
\nonumber\\
=\,&(-\tfrac 34 a^\prime b_7+\tfrac 12 a^\prime b_9+b^\prime b_5)\delta_\alpha^\beta G_\mu
+(-\tfrac 34 a^\prime b_8-\Ii a^\prime b_9+b^\prime b_6)\delta_\alpha^\beta H_\mu
\nonumber\\
&+(-a^\prime b_5+a^\prime b_7-a^\prime b_9+b^\prime b_7)\sigma_{\mu\nu\alpha}{}^\beta G^\nu
\nonumber\\
&+(-a^\prime b_6+a^\prime b_8+2\Ii a^\prime b_9+b^\prime b_8)\sigma_{\mu\nu\alpha}{}^\beta H^\nu
\nonumber\\
&+(-\tfrac 12 a^\prime+b^\prime)b_9\sigma^{\nu\rho}{}_\alpha{}^\beta F_{\nu\rho\mu}
\nonumber\\
\equiv\,&
\delta_\alpha^\beta (b^\prime_5 G_\mu+b^\prime_6 H_\mu)
+\sigma_{\mu\nu\alpha}{}^\beta (b^\prime_7 G^\nu+b^\prime_8 H^\nu)
+b^\prime_9\sigma^{\nu\rho}{}_\alpha{}^\beta F_{\nu\rho\mu}
\label{add4}\\
D_\alpha\5\lambda^{\prime\,\da}{}_\mu=\,&D_\alpha (\5b^\prime \5\lambda_\mu
-\5a^\prime \5\sigma_{\nu\mu}\5\lambda^\nu)^\da=0
\label{add5}
\end{align}
In appendix \ref{algebra} it is shown that $b_9$ must not be zero in order that $D_\alpha$ and $\5D_\da$ fulfill the standard supersymmetry algebra (on-shell and up to gauge transformations), and therefore one always can achieve $b_9=-1$ by redefining $\lambda_\mu$ if necessary (this redefinition is simply 
$\lambda^\prime_\mu= -\lambda_\mu/b_9$). Therefore henceforth with no loss of generality we shall presume $b_9=-1$, and thus we also require $b^\prime_9=-1$. The coefficients of
 $\sigma^{\nu\rho}{}_\alpha{}^\beta F_{\nu\rho\mu}$ in \eqref{add4} show that $b_9=b^\prime_9$ imposes
\begin{align}
b^\prime-\tfrac 12 a^\prime=1
\label{add6}
\end{align}
\eqref{add2} and \eqref{add6} are three equations for the four complex parameters $a$, $b$, $a^\prime$, $b^\prime$ and thus leave only one independent complex parameter which we choose to be $b$. This gives $a$, $a^\prime$, $b^\prime$ in terms of $b$ according to
\begin{align}
a=2(b-1),\qd a^\prime=\frac{2(1-b)}{4b-3},\qd b^\prime=\frac{3b-2}{4b-3},\qd b\in\mathbb{C}\backslash \{\tfrac 34\}
\label{add7}
\end{align}
The value $b=3/4$ is excluded because for this value the field redefinition 
$\lambda_\mu=b\lambda^\prime{}_\mu
+2(b-1)\lambda^{\prime\, \nu}\sigma_{\nu\mu}$ is not invertible. This can be understood, for example, using that the second equation in \eqref{add1} and the first equation in \eqref{add7} imply
\begin{align}
\left( \begin{array}{c} \lambda_\mu\\ \lambda^\nu\sigma_{\nu\mu}\end{array}\right)
=\left( \begin{array}{cc}b&2(b-1)\\ \tfrac 32(b-1)&3b-2\end{array}\right)
\left( \begin{array}{c} \lambda^\prime_\mu\\ \lambda^{\prime\,\nu}\sigma_{\nu\mu}\end{array}\right)
\label{add1a}
\end{align}
The invertibility of the matrix in \eqref{add1a} requires the non-vanishing of the determinant $4b-3$ of the matrix and thus $b\neq 3/4$.

As is also shown in appendix \ref{algebra} the standard supersymmetry algebra imposes for $b_9=-1$ that $b_2$ and $b_4$ fulfill $b_2=1-b_1$ and $b_4=\Ii(1-b_1)$. Using this and $a=2(b-1)$ in \eqref{add3} one obtains, in particular,
\begin{align}
\5b^\prime_1=b(4\5b_1-3)+3(1-\5b_1)
\label{add8}
\end{align}
Analogously to $4b-3\neq 0$ one has $4\5b_1-3\neq 0$, cf. equation \eqref{b28}.
Hence, \eqref{add8} shows that for every value of $b_1$ there is precisely one value of $b$ such that 
$b^\prime_1=1$:
\begin{align}
b=\frac{3\5b_1-2}{4\5b_1-3}\qd \Rightarrow\qd b^\prime_1=1
\label{add9}
\end{align}
In other words, in addition to $b_9=-1$ we can presume $b_1=1$ with no loss of generality. Therefore we shall consider these cases now. 
In appendix \ref{algebra} it is shown that in these cases the standard supersymmetry algebra 
particularly imposes $y_3(y_6-\tfrac 32 y_8)=0$ and thus either 
(i) $y_3=0$, or (ii) $y_3\neq 0$ and $y_6=\tfrac 32 y_8$.
The cases (i) lead to the parameters in \eqref{sol1} which provides the theories 
presented in section \ref{ex}. 

In the cases (ii) appendix \ref{algebra} gives
\begin{align}
b_1=1,\ b_3=y_3(y_5+\Ii)\neq 0,\ b_5=\Ii y_5,\ b_9=-1, \ b_2=b_4=b_6=b_7=b_8=0.
\label{add10}
\end{align}
The parameters $b_i$ in \eqref{add10} fulfill equations \eqref{b34} except for the parameter $b_3$.  Now, the only coefficient $c_i$ in the equations \eqref{c2} which depends on $b_3$ is the coefficient $c_2$. As a consequence none of the equations \eqref{11.a1} through \eqref{11.6} changes in the cases (ii) except for the equations \eqref{11.12} and \eqref{11.11}. It is easy to derive that the equations which replace equations \eqref{11.12} and \eqref{11.11} in the cases (ii) give $a_6=0$ and
$a_5-2a_8-3a_7y_5=0$ (whereas in the cases (i) equations \eqref{11.12} and \eqref{11.11} just give $a_6=0$).
The equations \eqref{11.10} through \eqref{11.6} additionally give $2a_8= a_5y_6$ and $a_7=-a_5y_5$, both in the cases (i) and in the cases (ii), with $y_6=0$ in the cases (ii). In the cases (ii) one obtains $a_5=a_6=a_7=a_8=0$ which implies that the free field Lagrangian \eqref{b5} is zero (in the cases (i) one obtains $a_6=0$, $a_7=-a_5y_5$, 
$a_8=\tfrac 12 a_5 y_6$ where $a_5$, $y_5$, $y_6$ are arbitrary real parameters). Hence, the cases (ii) do not provide a nontrivial supersymmetric free field theory, in contrast to the cases (i).

\mysection{Conclusion}\label{conc}

This paper shows that standard supergravity theories are not the only supersymmetric gauge theories of spacetime translations. Rather, in four spacetime dimensions there is another class of theories, presented in section \ref{ex}, which are \textit{new} (previously unknown) and similar to globally supersymmetric Yang-Mills theories in flat spacetime. In these theories, the supersymmetries are \textit{global} symmetries generated by infinitesimal supersymmetry transformations whose commutators contain infinitesimal general coordinate transformations with field dependent gauge parameters, analogously to the presence of infinitesimal Yang-Mills gauge transformations in the commutators of infinitesimal global supersymmetry transformations in supersymmetric Yang-Mills theories in flat spacetime. Furthermore it is shown that, up to local field redefinitions, this class of theories provides all 4D supersymmetric gauge theories of spacetime translations different from standard supergravity which are in the scope of investigation of the present paper (this scope particularly specifies the field content of the theories).\footnote{The results of the present paper do not exclude that there are further 4D supersymmetric gauge theories of spacetime translations when one allows an extended field content. In particular there may be 4D supersymmetric gauge theories of spacetime translations with extended ($N>1$) supersymmetry different from standard extended supergravity theories.} In particular this implies that supersymmetric versions of ``teleparallel'' theories with Lagrangians \eqref{a24} within this scope only exist for coefficients $a_1\dots,a_4$ fulfilling either (a) $a_2=a_3=a_4=0$ ($\Leftrightarrow$ $a^\prime_2=a_3=a^\prime_4=0$) or (b) $a_2=-2a_1$, $a_3=-4a_1$, $a_4=0$
($\Leftrightarrow$ $a^\prime_2=4a^\prime_1$,
$a_3=-2a^\prime_1$, $a^\prime_4=0$). Case (a) is represented by the theories in section \ref{ex}, case (b) by standard 4D $N=1$ supergravity.

\appendix

\mysection{Conventions}\label{conv}

Minkowski metric, $\epsilon$-symbols:
\begin{align}
&\eta_{ab}=\mathrm{diag}(1,-1,-1,-1),\qd a,b\in\{0,1,2,3\}
\nonumber\\
&\epsilon^{abcd}=\epsilon^{[abcd]},\qd \epsilon^{0123}=1\nonumber\\
&\epsilon^{\alpha\beta}=-\epsilon^{\beta\alpha},\ \alpha,\beta\in\{1,2\},\qd
\epsilon^{\da\dbe}=-\epsilon^{\dbe\da},\ \da,\dbe\in\{\dot 1,\dot 2\},\qd
\epsilon^{12}=\epsilon^{\dot 1\dot 2}=1\nonumber\\
&\epsilon_{\alpha\gamma}\epsilon^{\gamma\beta}=\delta_\alpha^\beta,\qd
\epsilon_{\da\dga}\epsilon^{\dga\dbe}=\delta_\da^\dbe \nonumber         
\end{align}
Matrices $\sigma^a$ with entries $\sigma^a{}_{\alpha\da}$
($\alpha$: row index, $\da$: column index):
\[
\sigma^0=\left( \begin{array}{rr}1&0\\ 0&1\end{array}\right),\qd
\sigma^1=\left( \begin{array}{rr}0&1\\ 1&0\end{array}\right),\qd
\sigma^2=\left( \begin{array}{rr}0&-\Ii\\ \Ii&0\end{array}\right),\qd
\sigma^3=\left( \begin{array}{rr}1&0\\ 0&-1\end{array}\right)  
\]
Matrices $\5\sigma^a$ with entries $\5\sigma^{a\da\alpha}$:
\begin{align} 
\5\sigma^{a\, \da\alpha}=
\epsilon^{\da\dbe}\epsilon^{\alpha\beta}\sigma^a_{\beta\dbe}    
\nonumber
\end{align}
Matrices $\sigma^{ab},\5\sigma^{ab}$:
\begin{align} 
\sigma^{ab}=
\tfrac 14(\sigma^a\5\sigma^b-\sigma^b\5\sigma^a),\qd
\5\sigma^{ab}
=\tfrac 14(\5\sigma^a\sigma^b-\5\sigma^b\sigma^a) 
\nonumber
\end{align}
Raising and lowering of spinor indices:
\begin{align} 
\psi_\alpha=\epsilon_{\alpha\beta}\psi^\beta,\qd
\psi^\alpha=\epsilon^{\alpha\beta}\psi_\beta,\qd
\5\psi_\da=\epsilon_{\da\dbe}\5\psi^\dbe,\qd
\5\psi^\da=\epsilon^{\da\dbe}\5\psi_\dbe
\nonumber
\end{align}
Contraction of spinor indices:
\begin{align}
 \psi\chi=\psi^\alpha\chi_\alpha,\qd
\5\psi\5\chi=\5\psi_\da\5\chi^\da
\nonumber
\end{align}
Symmetrization and antisymmetrization of indices are defined with ``weight one'', e.g.:
\begin{align}
X_{(ab)}=\tfrac 12(X_{ab}+X_{ba}),\qd
X_{[ab]}=\tfrac 12(X_{ab}-X_{ba})
\nonumber
\end{align}

\mysection{Derivation of equations \eqref{b11} and \eqref{b34}}\label{algebra}

\eqref{b1}, \eqref{b2} and the complex conjugates thereof give:
\begin{align}
z^{-1}\acom {D_\alpha}{\5D_\da}A_{\mu\nu}=\,&
(\tfrac 12b_7\5b_1+b_5\5b_2-\Ii b_7\5b_4+\Ii b_9\5b_4) G_\mu\sigma_{\nu\alpha\da}
\label{bb1}\\
&+(b_5\5b_1+\tfrac 12b_7\5b_2+\Ii b_7\5b_4-\Ii b_9\5b_4) G_\nu\sigma_{\mu\alpha\da}
\label{bb2}\\
&+(\tfrac 12b_8\5b_1+b_6\5b_2-\Ii b_8\5b_4+2b_9\5b_4) H_\mu\sigma_{\nu\alpha\da}
\label{bb3}\\
&+(b_6\5b_1+\tfrac 12b_8\5b_2+\Ii b_8\5b_4-2b_9\5b_4) H_\nu\sigma_{\mu\alpha\da}
\label{bb4}\\
&+b_9(\5b_1F_{\rho\mu\nu}+\5b_2F_{\rho\nu\mu}-\Ii \5b_4 F_{\mu\nu\rho})\sigma^\rho{}_{\alpha\da}
\label{bb5}\\
&
+\tfrac 12 [(2b_6-b_8)\5b_4+\Ii b_8(\5b_2-\5b_1)]H_{\mu\nu\rho}\sigma^\rho{}_{\alpha\da}
\label{bb5a}\\
&+[-\tfrac 12 b_7(\5b_1+\5b_2+3\5b_3)+(b_5+b_9)\5b_3]\eta_{\mu\nu}G_\rho\sigma^\rho{}_{\alpha\da}
\label{bb6}\\
&+[-\tfrac 12 b_8(\5b_1+\5b_2+3\5b_3)+b_6\5b_3]\eta_{\mu\nu}H_\rho\sigma^\rho{}_{\alpha\da}
\label{bb7}\\
&+\tfrac 12 b_9(\Ii \5b_1F^{\rho\sigma}{}_\nu\epsilon_{\mu\rho\sigma\tau}
+\Ii \5b_2F^{\rho\sigma}{}_\mu\epsilon_{\nu\rho\sigma\tau}
-\5b_4 F^{\rho\sigma}{}_\tau\epsilon_{\mu\nu\rho\sigma})
\sigma^\tau{}_{\alpha\da}
\label{bb8}\\
&+\tfrac 12 [(b_7-2b_5)\5b_4+\Ii b_7(\5b_1-\5b_2)]
\epsilon_{\mu\nu\rho\sigma} G^\rho\sigma^\sigma{}_{\alpha\da}+c.c.
\label{bb9}
\end{align}
where 
\begin{align}
H_{\mu\nu\rho}=\epsilon_{\mu\nu\rho\sigma}H^\sigma=
\6_\mu B_{\nu\rho}+\6_\nu B_{\rho\mu}+\6_\rho B_{\mu\nu}
\label{b8}
\end{align}
\eqref{bb5} and \eqref{bb5a} have to give the terms on the right hand side of \eqref{b7}. The parts of \eqref{bb5} and \eqref{bb5a} which are symmetric in $\mu,\nu$ show that $b_9$ must not vanish. By redefining $\lambda_\alpha{}^\mu$ one can thus fix $b_9$ to some particular non-zero value. Hence, with no loss of generality one may use
\begin{align}
b_9=-1
\label{b9}
\end{align}

Then the parts of the right hand side of \eqref{b7} and of \eqref{bb5} which are symmetric in $\mu,\nu$ give
\begin{align}
x_1+x_2=1
\label{b10}
\end{align}
and $X_{\mu\alpha\da}=2zA_{\nu\mu}\sigma^\nu{}_{\alpha\da}+k_{\mu\alpha\da}+k_{\mu\nu\alpha\da}x^\nu$ where $k_{\mu\alpha\da}$ and $k_{\mu\nu\alpha\da}=-k_{\nu\mu\alpha\da}$ are constants. The constant contribution $k_{\mu\alpha\da}$ to $X_{\mu\alpha\da}$ can be ignored because it does not contribute to \eqref{b7}.
The parts of the right hand side of \eqref{b7} and of \eqref{bb5} and \eqref{bb5a} which are antisymmetric in $\mu,\nu$ then give $k_{\mu\nu\alpha\da}=0$ which implies \eqref{b11}, and
\begin{align}
y_4=\,&1-x_1
\label{b12}\\
0=\,&(1-x_1)(2+y_6-\tfrac 12 y_8)
+x_4(x_6-\tfrac 12 x_8)
+\tfrac 12 y_8(2x_1-1)
-\tfrac 12 x_8(y_1-y_2)
\label{b13}
\end{align}
where \eqref{b12} arises from the terms containing derivatives of $H_{\mu\nu}$, \eqref{b13} arises from the terms containing derivatives of $B_{\mu\nu}$, and \eqref{b11}, \eqref{b9} and \eqref{b10} were used (and additionally \eqref{b12} to derive \eqref{b13}).

By \eqref{b7}, in addition to \eqref{b9} through
\eqref{b13}, the real parts of the coefficents in \eqref{bb1} through \eqref{bb4} and \eqref{bb6} through \eqref{bb9} have to vanish where the parts of \eqref{bb8} and \eqref{bb9} which are antisymmetric in $\mu,\nu$ must be considered together (as these parts both contain 
$\epsilon_{\mu\nu\rho\sigma} G^\rho\sigma^\sigma{}_{\alpha\da}$). We shall now work out these requirements using \eqref{b9} through \eqref{b12}. 
We start with \eqref{bb8}. The parts of \eqref{bb8} which are symmetric in $\mu,\nu$ give
\begin{align}
y_1+y_2=0
\label{b14}
\end{align}
and thus, together with \eqref{b10}:
\begin{align}
b_1+b_2=1
\label{b15}
\end{align}
The parts of \eqref{bb8} and \eqref{bb9} which are antisymmetric in $\mu,\nu$ give, using \eqref{b14}:
\begin{align}
x_4=\,&y_1
\label{b16}\\
0=\,&y_1(x_5-\tfrac 32 x_7-1)+y_5(1-x_1)+y_7(\tfrac 32 x_1-1)
\label{b17}
\end{align}
\eqref{b12} and \eqref{b16} give
\begin{align}
b_4=\Ii (1-b_1)
\label{b18}
\end{align}
Using \eqref{b9}, \eqref{b15} and \eqref{b18}, \eqref{bb1} and \eqref{bb2} give
\begin{align}
x_7=\,&-2x_5
\label{b19}\\
0=\,&x_5(4x_1-3)+y_1(y_5-\tfrac 32 y_7)+1-x_1
\label{b20}
\end{align}
Analogously \eqref{bb3} and \eqref{bb4} give
\begin{align}
x_8=\,&-2x_6
\label{b21}\\
0=\,&x_6(4x_1-3)+y_1(y_6-\tfrac 32y_8+2)
\label{b22}
\end{align}
Using \eqref{b9}, \eqref{b15}, \eqref{b18} and \eqref{b19}, \eqref{bb6} gives
\begin{align}
x_5-x_3(1-4x_5)+y_3(y_5-\tfrac 32 y_7)=0
\label{b23}
\end{align}
Finally, using \eqref{b15} and \eqref{b21}, \eqref{bb7} gives
\begin{align}
x_6(4x_3-1)+y_3(y_6-\tfrac 32 y_8)=0
\label{b24}
\end{align}
\eqref{b9} through \eqref{b24} still leave a lot of freedom for the coefficients $x_i$ and $y_i$. The choice $y_1=0$ reduces this freedom. For this choice \eqref{b16} and \eqref{b17} yield
\begin{align}
&y_1=x_4=0
\label{b26}\\
&y_5(1-x_1)=y_7(1-\tfrac 32 x_1)
\label{b27}
\end{align}
Using $y_1=0$ in \eqref{b20}, one observes that $x_1=3/4$ would give $0=1/4$. Hence, for $y_1=0$, $x_1$ must not be $3/4$:
\begin{align}
y_1=0:\qd x_1\neq \tfrac 34
\label{b28}
\end{align}
\eqref{b20} through \eqref{b22} now give
\begin{align}
x_5=\,&\frac{1-x_1}{3-4x_1}
\label{b29}\\
x_6=\,&x_8=0
\label{b30}
\end{align}
and,  using \eqref{b30}, \eqref{b13} gives
\begin{align}
(1-x_1)(2+y_6)=(1-\tfrac 32 x_1)y_8
\label{b31}
\end{align}
Using \eqref{b29} in \eqref{b23} gives
\begin{align}
x_3=x_1-1+(4b_1-3)y_3(y_5-\tfrac 32 y_7)
\label{b23a}
\end{align}
One is left with \eqref{b24} which gives, using \eqref{b30}:
\begin{align}
y_3(y_6-\tfrac 32 y_8)=0
\label{b32}
\end{align}
Hence, in the cases $y_1=0$ there are two options: $y_3=0$, or $y_3\neq 0$ and $y_6=\tfrac 32 y_8$. The first option gives \eqref{b34}. We note that the second option provides the same results for the coefficients $b_2, b_4, b_5,b_7$ as in \eqref{b34}, and 
$b_3=b_1-1+(4b_1-3)y_3(y_5-\tfrac 32 y_7)+\Ii y_3$, $b_6=6\Ii (b_1-1)$ and $b_8=4\Ii (b_1-1)$.

\mysection{Derivation of Lagrangians}\label{brst}

In this appendix it is explained how one can derive Lagrangians which are invariant, up to total divergences, under general coordinate transformations and supersymmetry transformations using BRST methods. The approach uses general and well established concepts and results, cf. e.g. \cite{Brandt:1996mh,Barnich:2000zw,Dragon:2012au}, 
and more specific results on supersymmetric theories \cite{Brandt:1992ts,Brandt:1993vd,Brandt:1996au}. We shall not review these concepts and results in detail. However, for readers which are not familiar with them we shall outline the main line of reasoning. 

We shall first discuss supersymmetric pure ``teleparallel''  theories whose fields are $\viel a\mu$, $\lambda^{\alpha a}$, $\5\lambda^{\da a}$ and auxiliary fields $B^a$ (in place of $B^a$ one may use $\7B^a$, cf. \eqref{b7a} and \eqref{b8a}, or other auxiliary fields -- if any). As in section \ref{tele} we shall presume that the commutator algebra of the infinitesimal supersymmetry transformations and general coordinate transformations closes off-shell. Then one can construct a BRST differential $s$ which acts on the fields and on ``ghosts'' related to infinitesimal general coordinate transformations and global supersymmetry transformations, and which squares to zero on all fields and the ghosts.\footnote{Actually $s$ is an ``extended'' BRST differential for local and global symmetries \cite{Brandt:1997cz}.} The BRST transformations of the 
fields are:
\begin{align}
s\viel a\mu=\,&C^\nu\6_\nu\viel a\mu+(\6_\mu C^\nu)\viel a\nu
+(\xi^\alpha\cD_\alpha+\5\xi^\da\5\cD_\da)\viel a\mu
\label{brs1}
\\
s\phi^M=\,&(C^\mu\6_\mu+\xi^\alpha\cD_\alpha+\5\xi^\da\5\cD_\da)\phi^M
\label{brs2}
\end{align}
where $C^\mu$ are anticommuting ghost fields of general coordinate transformations and $\xi^\alpha$ and $\5\xi^\da$ are commuting constant ghosts of global supersymmetry transformations, and 
\begin{align}
\{\phi^M\}=\{\lambda^{\alpha a},\5\lambda^{\da a},B^a\}
\label{brs3}
\end{align}
The BRST transformations of the ghosts are:
\begin{align}
s C^\mu=C^\nu\6_\nu C^\mu+2z\xi\sigma^\mu\5\xi,\qd 
s\xi^\alpha=s\5\xi^\da=0
\label{brs4}
\end{align}

Now, the BRST transformations of the fields are just infinitesimal general coordinate transformations and supersymmetry transformations with parameters of these transformations replaced by the respective ghosts (up to the factor $z$ in the definition of the supersymmetry transformations \eqref{d4}). Hence, a Lagrangian $L$ constructed of the fields and their derivatives is invariant, up to total divergences, under general coordinate transformations and supersymmetry transformations if and only if $sL$ is a total divergence. Using differential forms this can be written as
\begin{align}
s\omega_4+d\omega_3=0
\label{brs5}
\end{align}
where $\omega_4=d^4x L$ is a local 4-form with ghost number 0 and $\omega_3$ is a local 3-form with ghost number 1, where the ghost number is the degree of homogeneity in the ghosts, and $d$ denotes the exterior derivative
\begin{align}
d=dx^\mu\6_\mu
\label{brs6}
\end{align}
The differentials $dx^\mu$ are treated as anticommuting variables which are BRST-invariant ($sdx^\mu=0$). Using $\com{s}{\6_\mu}=0$, this gives
\begin{align}
s^2=\acom{s}{d}=d^2=0
\label{brs7}
\end{align}
Applying $s$ to \eqref{brs5} and using \eqref{brs7} gives $ds\omega_3=0$. The fact that the cohomology of $d$ is trivial in the space of local forms in form-degrees $0<p<4$ (i.e., $d\omega_p=0$ implies $\omega_p=d\eta_{p-1}$ for $0<p<4$ in four spacetime dimensions) implies $s\omega_3+d\omega_2=0$ for some local 2-form $\omega_2$ with ghost number 2. Repeating the reasoning one concludes $s\omega_2+d\omega_1=0$,
$s\omega_1+d\omega_0=0$ and $s\omega_0=0$ for some local 1-form $\omega_1$ with ghost number 3 and some local 0-form $\omega_0$ with ghost number 4 (a constant form with form-degree 0 cannot occur here because that form would be a polynomial in the supersymmetry ghosts which cannot arise as it would be independent of the fields). Hence, the forms $\omega_4,\ldots,\omega_0$ fulfill the so-called descent equations. These equations can be  written compactly as
\begin{align}
\4s\omega=0,\qd \4s=s+d,\qd \omega=\sum_{p=0}^4\omega_p
\label{brs8}
\end{align}

Sums of local forms, such as $\omega$, will be called ``total forms''. $\4s$ is a differential ($\4s^2=0$, as follows from \eqref{brs7}) which has ``total degree'' 1 where the total degree is the sum of the form degree and the ghost number, i.e. $\4s$ increases the total degree by one unit.
Hence, any Lagrangian which is invariant, up to total divergences, under general coordinate transformations and supersymmetry transformations gives rise to a local total form $\omega$ with total degree 4 which is $\4s$-closed. Furthermore one can presume that $\omega$ is not $\4s$-exact because $\omega= \4s\eta$ 
for some local total form $\eta$ with total degree 3 would imply $\omega_4=d\eta_3$ (with $\eta_3$ the 3-form in $\eta$), and thus that $L$ is a total divergence. Hence, $\omega$ is determined by the cohomology $H(\4s)$ of $\4s$ in the space of local total forms of the fields and ghosts at total degree 4.

Now, $H(\4s)$ can be analysed analogously as in standard supergravity in \cite{Brandt:1993vd,Brandt:1996au}. Firstly one introduces
appropriate variables that substitute for the fields, ghosts and derivatives thereof:
\begin{align}
&\{\cU^\ell\}=\{x^\mu,\6_{(\mu_1}\ldots\6_{\mu_k}\viel a{\mu_{k+1})}:k=0,1,\ldots\},
\label{brs9}\\
&\{\cV^\ell\}=\{\4s\,\cU^\ell\}=\{dx^\mu,\6_{\mu_1}\ldots\6_{\mu_{k+1}}\4\xi^a+\ldots:k=0,1,\ldots\}
\label{brs10}
\\
&\4\xi^a=(C^\mu+dx^\mu)\viel a\mu
\label{brs11}\\
&\{\cT^r\}=\{\cD_{(a_1}\ldots \cD_{a_k}\T{a_{k+1})a_{k+2}}b, 
\cD_{(a_1}\ldots \cD_{a_k)}\phi^M:k=0,1,\ldots
\}
\label{brs12}
\end{align}
This yields
\begin{align}
\4s\,\cT^r=\,&(\4\xi^a\cD_a+\xi^\alpha\cD_\alpha+\5\xi^\da\5\cD_\da)\cT^r
\label{brs13}
\\
\4s\,\4\xi^a=\,&
2z\xi\sigma^a\5\xi
-\4\xi^b\xi^\alpha\T{\alpha b}a
-\4\xi^b\5\xi^\da\T{\da b}a
+\tfrac 12\4\xi^b\4\xi^c\T{bc}a
\label{brs14}
\\
\4s\,\xi^\alpha=\,&\4s\,\5\xi^\da=0
\label{brs15}
\end{align}
with $\T{\alpha b}a$, $\T{\da b}a$ and $\T{bc}a$ as in section \ref{tele}.

\eqref{brs9} through \eqref{brs15} imply that the $\cU^\ell$ and $\cV^\ell$ drop out of the cohomology $H(\4s)$ because they form so-called contractible pairs. It follows that $H(\4s)$ reduces to the cohomology of $\4s$ in the space of total forms $\omega(\4\xi,\xi,\5\xi,\cT)$ depending only on the $\cT^r$, $\4\xi^a$, $\xi^\alpha$ and $\5\xi^\da$. Hence, up to a total divergence, any Lagrangian which is invariant, up to total divergences, under general coordinate transformations and supersymmetry transformations as in section \ref{tele} is the 4-form contained in a total form $\omega(\4\xi,\xi,\5\xi,\cT)$ with total degree 4 which fulfills
\begin{align}
\4s\omega(\4\xi,\xi,\5\xi,\cT)=0
\label{brs16}
\end{align}
In order to find solutions $\omega(\4\xi,\xi,\5\xi,\cT)$ of \eqref{brs16} we decompose \eqref{brs16} into parts with different degree of homogeneity in the $\4\xi^a$ (this degree will be called $\4\xi$-degree in the following). According to \eqref{brs13} through \eqref{brs15}, $\4s$ decomposes 
on the $\cT^r$, $\4\xi^a$, $\xi^\alpha$, $\5\xi^\da$ according to
$\4s=\delta_-+\delta_0+\delta_+$ into three parts $\delta_-$, $\delta_0$ and $\delta_+$ with $\4\xi$-degrees $-1$, 0 and $+1$, respectively:
\begin{align}
&\delta_-=2z\xi\sigma^a\5\xi\, \frac{\6}{\6\4\xi^a}
\label{brs17}\\
&\delta_0=-(\4\xi^b\xi^\alpha\T{\alpha b}a+\4\xi^b\5\xi^\da\T{\da b}a)\, \frac{\6}{\6\4\xi^a}
+(\xi^\alpha\cD_\alpha\cT^r+\5\xi^\da\5\cD_\da \cT^r)\,\frac {\6}{\6\cT^r}
\label{brs18}\\
&\delta_+=
\tfrac 12\4\xi^b\4\xi^c\T{bc}a\, \frac{\6}{\6\4\xi^a}+(\4\xi^a\cD_a\cT^r)\,\frac {\6}{\6\cT^r}
\label{brs19}
\end{align}
The decomposition of \eqref{brs16} thus gives
\begin{align}
\delta_-\omega^{\ul{m}}=0,\ \delta_-\omega^{\ul{m}+1}+\delta_0\omega^{\ul{m}}=0,\ \ldots
\label{brs20}
\end{align}
where $\omega^k$ denotes the part of $\omega$ with $\4\xi$-degree $k$, and $\ul{m}$ denotes the lowest $\4\xi$-degree in this decomposition,
\begin{align}
\omega(\4\xi,\xi,\5\xi,\cT)=\sum_{k=\ul{m}}^4\omega^k,\qd 
\4\xi^a\,\frac{\6\omega^k}{\6\4\xi^a}=k\omega^k
\label{brs21}
\end{align}
By \eqref{brs20} $\omega^{\ul{m}}$ is $\delta_-$-closed. Furthermore, one can presume that $\omega^{\ul{m}}$ is not $\delta_-$-exact and does not contain any $\delta_-$-exact portion $\delta_-\eta^{\ul{m}+1}$ because such a portion can be removed from $\omega$ by subtracting $\4s\eta^{\ul{m}+1}$ from $\omega$ (which would alter $L$ at most by a total divergence).
Hence, $\omega^{\ul{m}}$ is determined by the cohomology $H(\delta_-)$ of $\delta_-$.
As $\delta_-$ only involves the ghosts $\4\xi^a$, $\xi^\alpha$, $\5\xi^\da$, the cohomology $H(\delta_-)$ can be formulated on polynomials 
$f(\4\xi,\xi,\5\xi)$ of these ghosts. According to \cite{Brandt:1992ts} one has
\begin{align}
&\delta_-f(\4\xi,\xi,\5\xi)=0\ \Leftrightarrow\ 
f(\4\xi,\xi,\5\xi)=P(\5\vartheta,\xi)+P^\prime(\vartheta,\5\xi)+r\Theta 
+\delta_-g(\4\xi,\xi,\5\xi)
\label{brs22}\\
&P(\5\vartheta,\xi)+P^\prime(\vartheta,\5\xi)+r\Theta 
=\delta_-g(\4\xi,\xi,\5\xi)\ \Leftrightarrow\ 
P+P^\prime=0,\ r=0
\label{brs23}
\end{align}
where
\bea
\vartheta^\alpha=\5\xi_\da\4\xi^{\da\alpha},\ 
\5\vartheta^\da=\4\xi^{\da\alpha}\xi_\alpha,\ 
\Theta=\5\xi_\da\4\xi^{\da\alpha}\xi_\alpha
\label{brs24}
\eea
with $\4\xi^{\da\alpha}=\4\xi_a\5\sigma^{a\da\alpha}$, and in \eqref{brs23} $P+P^\prime=0$ can occur only in the trivial case that $P$ and $P^\prime$ do not depend on the ghosts at all. \eqref{brs22} through \eqref{brs24} state that $H(\delta_-)$ is represented by polynomials $P(\5\vartheta,\xi)$ in the $\5\vartheta^\da$ and $\xi^\alpha$, 
polynomials $P^\prime(\vartheta,\5\xi)$ in the $\vartheta^\alpha$ and $\5\xi^\da$, and a representative proportional to $\Theta$.

As the $\5\vartheta^\da$ anticommute, $P(\5\vartheta,\xi)$ is a most bilinear in the $\5\vartheta^\da$, and any contribution to $P(\5\vartheta,\xi)$ which is bilinear in the $\5\vartheta^\da$ reads $\5\vartheta\5\vartheta Q(\xi)$ for some
polynomial $Q(\xi)$ in the $\xi^\alpha$. Hence, any polynomial $P(\5\vartheta,\xi)$ only contains terms which have at most $\4\xi$-degree 2. Analogous statements apply to $P^\prime(\vartheta,\5\xi)$. Furthermore $\Theta$ has total degree 3.
This implies that the part $\omega^{\ul{m}}$ of any nontrivial real $\4s$-cocycle $\omega(\4\xi,\xi,\5\xi,\cT)$ with total degree 4 can be presumed to have $\4\xi$-degree $\ul{m}\in\{\ul{0},\ul{1},\ul{2}\}$ and in the various cases can be written as:
\begin{align}
\ul{m}=\ul{2}:&\qd \omega^{\ul{2}}=\5\vartheta\5\vartheta h(\cT)+c.c.
\label{brs25}\\
\ul{m}=\ul{1}:&\qd \omega^{\ul{1}}=\5\vartheta^\da\xi^\alpha\xi^\beta h_{\da\alpha\beta}(\cT)+c.c.
\label{brs26}\\
\ul{m}=\ul{0}:&\qd \omega^{\ul{0}}=\xi^\alpha\xi^\beta\xi^\gamma\xi^\delta  h_{\alpha\beta\gamma\delta}(\cT)+c.c.
\label{brs27}
\end{align}

Now one concludes that any $\4s$-cocycle $\omega(\4\xi,\xi,\5\xi,\cT)$  containing a Lagrangian with terms given in \eqref{a24} has a part $\omega^{\ul{2}}$ as in \eqref{brs25} with $h(\cT)$ quadratic in the $\lambda^{\alpha a}$ and/or $\5\lambda^{\da a}$. To show this we assign the following mass dimensions to the ghosts and differentials:
\begin{align}
[C^\mu]=[dx^\mu]=-1,\ [\xi^\alpha]=[\5\xi^\da]=-1/2
\label{brs28}
\end{align}
\eqref{d9} and \eqref{brs28} imply that both $s$ and $d$ have mass dimension 0. As $d^4x L_\mathrm{Bose}$ with $L_\mathrm{Bose}$ as in \eqref{a24} has mass dimension $-2$, a corresponding $\4s$-cocycle $\omega(\4\xi,\xi,\5\xi,\cT)$ also has mass dimension $-2$. Furthermore, this $\4s$-cocycle must be at least quadratic in the $\cT^r$. Now,  
$\5\vartheta\5\vartheta$, $\5\vartheta^\da\xi^\alpha\xi^\beta$ and $\xi^\alpha\xi^\beta\xi^\gamma\xi^\delta$ in \eqref{brs25} through \eqref{brs27} have
mass dimensions $-3$, $-5/2$ and $-2$, respectively. All the tensors $\cT^r$ in \eqref{brs12} have mass dimensions
$\geq 1/2$, where only $\lambda^{\alpha a}$ and $\5\lambda^{\da a}$ have mass dimension $1/2$. This implies that any $\4s$-cocycle $\omega(\4\xi,\xi,\5\xi,\cT)$  containing a Lagrangian with terms given in \eqref{a24} has an
$\omega^{\ul{2}}$ as in \eqref{brs25} with $h(\cT)$ quadratic in the $\lambda^{\alpha a}$ and/or $\5\lambda^{\da a}$. Therefore we now shall discuss \eqref{brs16} for $\omega$ with $\omega^{\ul{2}}$
as in \eqref{brs25}, but for general $h(\cT)$. The second equation in \eqref{brs20} imposes that $\delta_0\omega^{\ul{2}}$ is $\delta_-$-exact. Using \eqref{brs18} with $\T {\alpha a}b$ 
obtained from \eqref{d7} and \eqref{d10}, one gets
\begin{align}
\delta_0[\5\vartheta\5\vartheta h(\cT)]=
\5\vartheta\5\vartheta \xi^\alpha[(4b_1-3)(\sigma_a\5\lambda^a)_\alpha+\cD_\alpha]h(\cT)+\ldots
\label{brs29}
\end{align}
where ellipses denote terms depending on components of both $\xi$ and $\5\xi$. Using \eqref{brs23} one concludes that 
$\delta_0[\5\vartheta\5\vartheta h(\cT)+c.c.]$ is $\delta_-$-exact if and only if $h(\cT)$ fulfills \eqref{d21}.
The other equations in \eqref{brs20} do not impose any further condition because $H(\delta_-)$ is trivial at $\4\xi$-degrees 3 and 4. One obtains
\begin{align}
\omega^3=\,&\Ii z^{-1}\Xi_a[(2b_1-3)\lambda^a\xi+2(1-2b_1)\lambda_b\sigma^{ab}\xi+\xi\sigma^a\5\cD]h(\cT)
+c.c.
\label{brs31}\\
\omega^4=\,&-\Ii z^{-2}\Xi [\tfrac 14 \5\cD^2+\tfrac 12 (4b_1-3)\5\chi\5\cD
+\tfrac 18 (4b_1-3)^2\5\chi\5\chi+\tfrac 14\psi\psi]h(\cT)+c.c.
\label{brs32}
\end{align}
where
\begin{align}
\Xi=-\tfrac 1{24}\epsilon_{abcd}\4\xi^d\4\xi^c\4\xi^b\4\xi^a,\qd 
\Xi_a=-\tfrac 1{6}\epsilon_{abcd}\4\xi^d\4\xi^c\4\xi^b
\label{brs33}
\end{align}
The 4-form contained in \eqref{brs32} provides \eqref{d20}, up to a factor $\Ii/4$
(which may be absorbed by redefining $h$).

The above analysis can be extended to theories with super Yang-Mills multiplets and/or scalar multiplets. We shall now briefly outline this extension, presuming again that the commutator algebra of the infinitesimal supersymmetry transformations, general coordinate transformations and Yang-Mills gauge transformations closes off-shell. The BRST transformation \eqref{brs1} of the tetrad remains unchanged, and the BRST transformations of the other fields are
\begin{align}
sA_\mu{}^i=\,&C^\nu\6_\nu A_\mu{}^i+(\6_\mu C^\nu)A_\nu{}^i
+\6_\mu C^i+\f jki C^k A_\mu{}^j
+(\xi^\alpha\cD_\alpha+\5\xi^\da\5\cD_\da)A_\mu{}^i
\label{brs34}
\\
s\phi^M=\,&(C^\mu\6_\mu+C^i\delta_i+\xi^\alpha\cD_\alpha+\5\xi^\da\5\cD_\da)\phi^M
\label{brs35}
\end{align}
where the $C^i$ are anticommuting ghost fields of Yang-Mills transformations, 
\begin{align}
\{\phi^M\}=\{\lambda^{\alpha a},\5\lambda^{\da a},B^a,\lambda^{\alpha i},\5\lambda^{\da i},D^i,\varphi^m,\eta^{\alpha m},F^m,\5\varphi^{\5m},\5\eta^{\da \5m},\5F^{\5m}\},
\label{brs36}
\end{align}
and $\delta_i\lambda^{\alpha j}=-\f ikj\lambda^{\alpha k}$, $\delta_i\varphi^m=-T_i{}^m{}_n\varphi^n$ etc. (again, $B^a$ in \eqref{brs36} may be replaced by other auxiliary fields). The BRST transformations of the ghosts $C^\mu$, $\xi^\alpha$ and $\5\xi^\da$ are as in \eqref{brs4}, and the BRST transformations of the Yang-Mills ghosts are
\begin{align}
sC^i=\,& C^\mu\6_\mu C^i+\tfrac 12 \f jki C^kC^j-2z\xi\sigma^\mu\5\xi A_\mu{}^i
\label{brs37}
\end{align}
The set $\{\cU^\ell\}$ of \eqref{brs9} now additionally contains the
$\6_{(\mu_1}\ldots\6_{\mu_k}A_{\mu_{k+1})}{}^i$, the set $\{\cV^\ell\}$ of \eqref{brs10} additionally 
contains the $\6_{\mu_1}\ldots\6_{\mu_{k+1}}\4C^i+\ldots$ (with $\4C^i$ as in \eqref{brs39}), and the set $\{\cT^r\}$ of \eqref{brs12}
now reads
\begin{align}
\{\cT^r\}=\{&\cD_{(a_1}\ldots \cD_{a_k}\T{a_{k+1})a_{k+2}}b,
 \cD_{(a_1}\ldots \cD_{a_k}\F{a_{k+1})a_{k+2}}i,
 \nonumber\\
&\cD_{(a_1}\ldots \cD_{a_k)}\phi^M:k=0,1,\ldots
\}
\label{brs38}
\end{align}
with the $\phi^M$ of \eqref{brs36}. The undifferentiated Yang-Mills ghosts give rise to additional variables
$\4C^i$ defined according to
\begin{align}
\4C^i=C^i+(C^\mu+dx^\mu)A_\mu{}^i
\label{brs39}
\end{align}
The $\4s$-transformation of $\4C^i$ is
\begin{align}
\4s\,\4C^i=
\tfrac 12 \f jki \4C^k\4C^j
-\4\xi^a\xi\sigma_a\5\lambda^i
-\4\xi^a\lambda^i\sigma_a\5\xi
+\tfrac 12\4\xi^a\4\xi^b\F{ab}i
\label{brs40}
\end{align}
In place of \eqref{brs13} one gets
\begin{align}
\4s\,\cT^r=\,&(\4\xi^a\cD_a+\4C^i\delta_i+\xi^\alpha\cD_\alpha+\5\xi^\da\5\cD_\da)\cT^r
\label{brs41}
\end{align}
\eqref{brs14} and \eqref{brs15} still hold unchanged.
Again the $\cU^\ell$ and $\cV^\ell$ drop out of the cohomology $H(\4s)$ because they form contractible pairs. Therefore
$H(\4s)$ reduces to the cohomology of $\4s$ in the space of total forms $\omega(\4\xi,\xi,\5\xi,\4C,\cT)$ depending only on the $\cT^r$, $\4\xi^a$, $\xi^\alpha$, $\5\xi^\da$ and $\4C^i$. The presence of the $\4C^i$ amends the structure of  $H(\4s)$ as compared to the case without super Yang-Mills multiplets. These amendments can be derived analogously to the analysis of $H(\4s)$ in \cite{Brandt:1996au}. We shall not discuss these amendments in detail here because they hardly are relevant to the derivation of Lagrangians. We only note that
the presence of $\4C^i\delta_i\cT^r$ in \eqref{brs41} imposes that any function $h(\cT)$ in \eqref{brs29} through \eqref{brs32} must be annihilated by the $\delta_i$, i.e.
\begin{align}
\forall i:\
\delta_i h(\cT)=0
\label{brs42}
\end{align}
One obtains that a Lagrangian which is invariant, up to total divergences, under general coordinate transformations, global supersymmetry transformations and Yang-Mills gauge transformations arises from \eqref{d20} 
for any function $h(\cT)$ which fulfills \eqref{d21} and \eqref{brs42}.

\end{document}